\documentclass[
reprint,
superscriptaddress,
 amsmath,amssymb,
aps,
pra,
]{revtex4-2}
\usepackage{accents}
\usepackage{graphicx}
\usepackage{dcolumn}
\usepackage{bm}
\usepackage{booktabs}
\usepackage{xcolor}
\usepackage{graphicx} 

\usepackage{soul} 
\usepackage{braket}
\usepackage{amsmath}

\newcommand{\doubleoverline}[1]{%
  \smash{\raisebox{-0.15ex}{\scalebox{0.9}{$\overline{\overline{#1}}$}}}%
}

\begin{document}
\preprint{APS/123-QED}
\title{Variational optical phase learning on a continuous-variable quantum compiler}
\thanks{This manuscript has been authored, in part, by UT-Battelle LLC, under
contract DE-AC05-00OR22725 with the U.S. Department of Energy (DOE). The publisher
acknowledges the U.S. government license to provide public access under the DOE Public
Access Plan (http://energy.gov/downloads/doe-public-access-plan).}

\author{Matthew A. Feldman}
\email{feldmanma@ornl.gov}
\affiliation{Quantum Information Science Section, Computational Sciences and Engineering Division, Oak Ridge National Laboratory, Oak Ridge, TN}
\affiliation{Quantum Science Center, Oak Ridge National Laboratory, Oak Ridge, TN, USA}
\author{Tyler Volkoff}%
\email{volkoff@lanl.gov}
\affiliation{Theoretical Division, Los Alamos National Laboratory, Los Alamos, NM, USA}%
\author{Seongjin Hong}
\affiliation{Quantum Science Center, Oak Ridge National Laboratory, Oak Ridge, TN, USA}
\author{Claire E. Marvinney}
\affiliation{Quantum Information Science Section, Computational Sciences and Engineering Division, Oak Ridge National Laboratory, Oak Ridge, TN}
\affiliation{Quantum Science Center, Oak Ridge National Laboratory, Oak Ridge, TN, USA}
\author{Zo\"{e} Holmes}
\affiliation{Laboratory of Quantum Information and Computation, Ecole Polytechnique Fédérale de Lausanne, Lausanne, Switzerland}
\author{Raphael C. Pooser}
\affiliation{Quantum Information Science Section, Computational Sciences and Engineering Division, Oak Ridge National Laboratory, Oak Ridge, TN}
\affiliation{Quantum Science Center, Oak Ridge National Laboratory, Oak Ridge, TN, USA}
\author{Andrew Sornborger}
\affiliation{Information Sciences, Los Alamos National Laboratory, Los Alamos, NM, USA }
\affiliation{Quantum Science Center, Oak Ridge National Laboratory, Oak Ridge, TN, USA}
\author{Alberto M. Marino}
\email{marinoa@ornl.gov}
\affiliation{Quantum Information Science Section, Computational Sciences and Engineering Division, Oak Ridge National Laboratory, Oak Ridge, TN}
\affiliation{Quantum Science Center, Oak Ridge National Laboratory, Oak Ridge, TN, USA}

\date{\today}

\begin{abstract}
Quantum process learning is a fundamental primitive that draws inspiration from machine learning with the goal of better studying the dynamics of quantum systems. One approach to quantum process learning is quantum compilation, whereby an analog quantum operation is digitized by compiling it into a series of basic gates. While there has been significant focus on quantum compiling for discrete-variable systems, the continuous-variable (CV) framework has received comparatively less attention. We present an experimental implementation of a CV quantum compiler that uses two mode-squeezed light to learn a Gaussian unitary operation. We demonstrate the compiler by learning a parameterized linear phase unitary through the use of target and control phase unitaries to demonstrate a factor of 5.4 increase in the precision of the phase estimation and a 3.6-fold acceleration in the time-to-solution metric when leveraging quantum resources. Our results are enabled by the tunable control of our cost landscape via variable squeezing, thus providing a critical framework to simultaneously increase precision and reduce time-to-solution.
\end{abstract}

\maketitle

\section{\label{sec:level1}Introduction}

There is growing awareness of the power of machine learning to process data from quantum experiments. However, to channel the spirit of Richard Feynman, if the underlying systems are inherently \textit{quantum}, perhaps \textit{quantum} machine learning is a more suitable general framework \cite{schuldIntroductionQuantumMachine,zhangRecentAdvancesQuantum}. Branching from fundamental results that extended the framework of neural networks to the discrete-variable (DV) quantum \cite{khatri2019quantum,heya2018variational,poland2020no, sharma2020reformulation} and continuous-variable (CV) quantum settings \cite{PhysRevResearch.1.033063}, this line of thought motivates the burgeoning field of quantum process learning \cite{cirstoiu2020variational,gibbs2021long, gibbs2022dynamical, geller2021experimental, huang2021quantum,caro2021generalization,caro2022outofdistribution, huang2022learning, caro2022learning}, where quantum training data are used to learn a quantum model for an unknown quantum process. Although quantum learning algorithms, unlike quantum process tomography or quantum sensing algorithms, rely on quantum training data, techniques from these approaches can often be applied in complementary ways toward characterizing quantum dynamics \cite{jerbi2023power,haiUniversalCompilationQuantum2022,haugGeneralizationQuantumGeometry2023,PhysRevLett.123.260505,PRXQuantum.4.020333,PhysRevLett.129.190501,mele2024learningquantumstatescontinuous}.

One key application of quantum process learning is quantum compilation, whereby a complex, potentially analogue, quantum operation is digitized by compiling it into a series of basic gates that are implementable on the available hardware forming a digital quantum twin. Such an approach can be used with the aim of learning shorter and more noise resilient circuits to reduce the resources required to implement large scale quantum algorithms. Quantum compilation has been proposed for quantum state tomography \cite{haiUniversalCompilationQuantum2022}, quantum entangled state preparation \cite{haiVariationalPreparationEntangled}, quantum circuit optimization \cite{Cincio_2018}, and to learn quantum dynamics~\cite{
tylerUniversalCompilingNo2021, volkoff2024learning,volkoffStrategiesVariationalQuantum2022,Jones2022robustquantum}. 

While receiving less attention, CV platforms provide fertile ground for almost all quantum-based applications, spanning from quantum computation~\cite{madsenQuantumComputationalAdvantage2022} and quantum machine learning~\cite{PhysRevResearch.1.033063, schuldQuantumMachineLearning2019} to quantum sensing~\cite{lawrieQuantumSensingSqueezed2019a, huaQuantumEnhancedProbes2023, guoDistributedQuantumSensing2020a}. The allure of these systems is rooted in their operational viability at ambient temperatures, their capacity for deterministic quantum entanglement \cite{andersen,furu}, and the effective management of quantum noise through the implementation of quantum error correcting codes~\cite{PhysRevA.64.012310,PhysRevX.10.011058,albertPerformanceStructureSinglemode2018, volkoffNonclassicalPropertiesQuantum2015}. Strides have been made in the integration of such algorithms within CV quantum optical systems, notably for tasks like variational eigenvalue determination~\cite{lauQuantumMachineLearning2017}.
Variational quantum algorithms have recently been implemented in the CV domain using single-mode squeezed light to optimize quantum sensing of an optical phase~\cite{nielsenVariationalQuantumAlgorithm2023} and a measurement-based quantum alternative operator ansatz (QAOA) algorithm using a non-variational quantum circuit~\cite{enomotoContinuousvariableQuantumApproximate2023}.

Recent theoretical work has shown that CV quantum processes can be learned using a two-mode squeezed state (TMSS) as a quantum entanglement resource~\cite{tylerUniversalCompilingNo2021,PhysRevLett.133.230604}. Two key features are made possible by the use of two-mode squeezing: 1) Only a single input-output pair is necessary for learning a single-mode unitary operation, and 2) the tunability of the degree of squeezing in a TMSS provides a mechanism to adaptively narrow or broaden the minimum of the cost landscape to aid training. Employing this method enables optimization of all parameters of a CV unitary operation, thus overcoming limitations of compiling in smaller dimensional subspaces, which may include reduced expressiveness and a constrained capacity for accurate representation of complex unitary transformations~\cite{arrazolaMachineLearningMethod2019}.  

In this work we demonstrate the first experimental implementation of a squeezing-based CV variational quantum compiler capable of learning a target unitary parameter. We employ a TMSS as a resource state to variationally learn the phase of a linear optical phase gate. Our quantum compilation algorithm (QCA) study demonstrates that quantum states of light can be used to control the width of gorges containing cost minima. We demonstrate that at low squeezing, the QCA cost function exhibits a larger gradient farther from the minimum, enabling the QCA to more easily find the minimum about the phase of interest. As the squeezing increases, the gradient near the minimum becomes steeper, allowing the algorithm to train more quickly and learn the phase parameter to a higher degree of precision.

\begin{figure*}[t!]
\includegraphics[scale=.5]{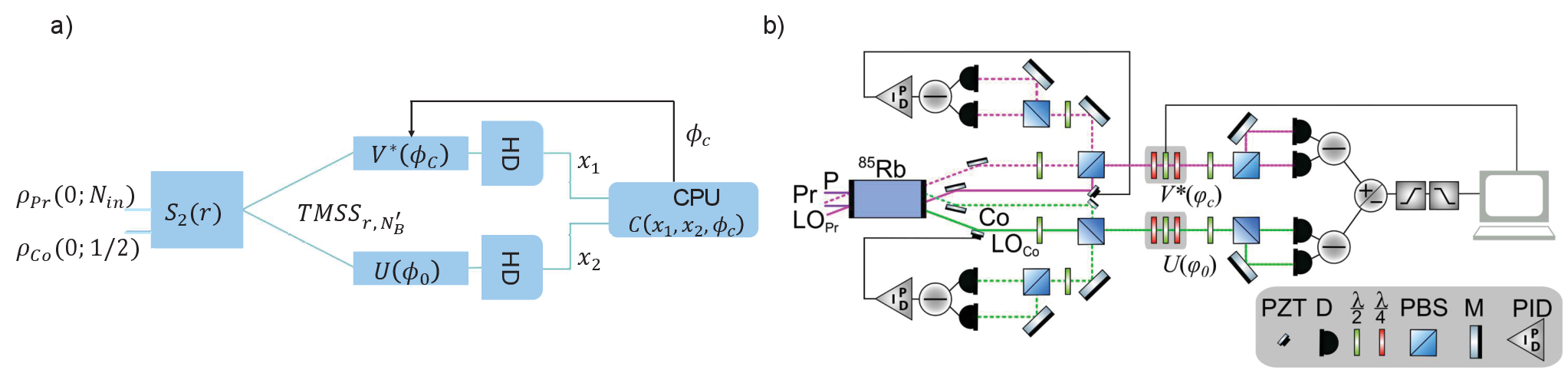}
\caption{\label{fig:1} (a) Schematic of a continuous-variable (CV) quantum compiler. A parameteric amplifier is weakly seeded with a noisy probe beam modeled as a thermal state, $\rho_{\text{Pr}}$, with no displacement, i.e. $\rho_{\text{Pr}}(0; N_{in}) = D(0) \rho_{\text{Pr}}(N_{in}) D^\dagger(0)$ where $D(\alpha)$ is the displacement operator, and $N_{in}$ is the noise of the probe. The modes of the TMSS interact with either the target, $U(\phi_0)$, or the control, $V^*(\phi_c)$, unitaries and are subsequently measured with homodyne detectors. These measurements are used to evaluate the cost function $C$, which is iteratively minimized using classical computing resources by adjusting the phase of the control unitary. (b) Experimental setup based on a truncated SU(1,1) interferometer for the implementation of the CV quantum compiler. 
Pr: probe; Rb: Rubidium vapor cell; Co: conjugate; $\text{LO}_{\text{Pr}}$: local oscillator for the probe;  $\text{LO}_{\text{Co}}$: local oscillator for the conjugate; PZT: piezoelectric transducer; D: detector; $\lambda/2$: half-wave plate; $\lambda/4$: quarter-wave plate; PBS: polarizing beam splitter; M: mirror; PID: proportional, integral, differential control. 
}
\end{figure*}

\section{\label{sec:level1}Methods}

\subsection{\label{sec:level2overview}QCA Overview}

We implement a CV QCA to  learn the phase of an optical phase gate. To this end, we employ a spatially truncated version of the quantum compiler suggested in Ref.~\cite{tylerUniversalCompilingNo2021}, as shown schematically in Fig.~\ref{fig:1}(a). Here, a noisy TMSS, with squeezing parameter $r$, is used as the quantum resource for the compilation. Such a noisy state allows us to take into account experimental imperfections and is obtained by taking one of the input modes (labeled probe) to be a thermal state with noise $\langle \Delta x_1^2 \rangle = \langle \Delta y_1^2 \rangle = N_{in}$ for the amplitude quadrature, $q_1 = (a  + a^\dagger)/ \sqrt{2}$ with eigenvalue $x_1$, and for the phase quadrature, $p_1 = i(a^\dagger-a)/ \sqrt{2}$ with eigenvalue $y_1$. The other input mode (labeled conjugate) is taken to be in the vacuum, such that it is shot noise limited. The modes of the TMSS interacts with either a target, $U(\phi_0)$, or a control, $V^*(\phi_c)$, optical phase unitary with phases $\phi_0$ and $\phi_c$, respectively, before being measured.  Given the focus on phase gates, a phase sensitive measurement, such as homoydne detection (HD) is needed to measure the quadratures. The measurement results are then fed to a classical processor to evaluate a cost function, $C$, that needs to be minimized to perform the compilation.

During the QCA, the parameter of the target unitary, $\phi_0$, is held constant. Through successive iterations the QCA evaluates the cost function and performs a gradient descent to update the phase, $\phi_c$, of the control unitary until the cost function has been minimized, at which point the QCA has learned the arbitrary phase shift, $\phi_0$, of the target unitary, $U(\phi_0)$. As we will see, the squeezing parameter, $r$, can be tuned to speed up the quantum compilation and increase the precision with which the parameter of the target unitary can be estimated.

\subsection{\label{sec:level2theory}QCA Cost Function Structure}

For the QCA to be successful, a proper cost function that depends on $\phi_{0}$ and $\phi_{c}$ must be identified.  Because our experiment measures homodyne data, we will define a cost function by considering the propagation of the Wigner function for a vacuum TMSS, $W(x_{1},y_{1},x_{2},y_{2})$,  through the system shown in Fig.~\ref{fig:1}(a). We consider the two modes of the TMSS, with respective ladder operators, ($a$, $a^{\dagger}$) and ($b$, $b^{\dagger}$), propagating through the corresponding phase gates, which introduce the local optical phase shift $e^{i\phi_{0}a^{\dagger}a}\otimes e^{-i\phi_{c}b^{\dagger}b}$. We model the seeding of the TMSS as a noisy state with input noise $N_{in} = N_B^{\prime} + 1/2$, where the variance $N_B^{\prime}$ designates the excess noise of the seed in our experiment. After propagation through these elements, the state can be described as a Gaussian state with zero mean and a covariance matrix of the form
\begin{align}
\Sigma &= {1+2\varepsilon N(1+N_{B}') \over 2}I_{2} \oplus {1+2\varepsilon ' (N(1+N_{B}')+N_{B}') \over 2}I_{2}  \nonumber \\
&{} +\sqrt{\varepsilon\varepsilon '}\sqrt{N(N+1)}(1+N_{B}')\cos\Delta \phi \, X\otimes Z \nonumber \\
&{} +\sqrt{\varepsilon\varepsilon '}\sqrt{N(N+1)}(1+N_{B}')\sin\Delta \phi \, X\otimes X,
\label{eqn:covv}
\end{align}
where the quadratures are ordered $(q_{1},p_{1},q_{2},p_{2})$ so that the correlations between the conjugate and probe modes are in the off-diagonal blocks, $X$ and $Z$ are Pauli matrices, $\Delta \phi := \phi_c - \phi_0$ is the phase difference between the control and target phase, $N=\sinh^{2}r$ is the photon intensity of each of the modes of the vacuum TMSS, and the transmissivity parameters for the conjugate and probe are $\varepsilon$ and $\varepsilon^{\prime}$, respectively. Equation~(\ref{eqn:covv}) can then be used to compute the Wigner function via the Fourier transform of the characteristic function \cite{serafini}. 

Given the correlations that are present between the amplitude difference and phase sum quadratures in a TMSS, to obtain a cost function that can be efficiently evaluated through homodyne detection, we choose to project the Wigner function onto the amplitude difference quadrature, $(q_{1}-q_{2})/\sqrt{2}$. The distribution of such joint quadrature with eigenvalue $X_{-} =(x_{1}-x_{2})/\sqrt{2}$  is obtained via 
\begin{equation}
    \rho(X_{-},r,\varepsilon,\varepsilon^{\prime}, N^{\prime}_{B},\Delta \phi) := {1\over 4}\int_{\mathbb{R}^{3}} dX_{+}dy_{1}dy_{2} W(x_{1},y_{1},x_{2},y_{2}),
    \label{eqn:uuu}
\end{equation}
where $y_{1}$ and $y_{2}$ are the eigenvalues of the phase quadratures of the respective modes and $X_{+}=(x_{1}+x_{2})/\sqrt{2}$ is the eigenvalue of the quadrature sum, $(q_{1}+q_{2})/\sqrt{2}$. 

Finally, to obtain a cost function that depends on $\Delta \phi$ and not on the measured eigenvalue, $X_{-}$, one can consider the zero'th order Taylor series expansion of $\rho(X_{-},r,\varepsilon,\varepsilon^{\prime}, N^{\prime}_{B},\Delta \phi)$ in a small neighborhood around $X_{-}=0$. We can then carry out the Gaussian integral in Eq.~(\ref{eqn:uuu}) to obtain the cost function

\begin{align}
C(r, \varepsilon, \varepsilon^{\prime}, N^\prime_B, \Delta \phi)
&:= -\rho(0, r, \varepsilon, \varepsilon^{\prime}, N_B^{\prime}, \Delta \phi)\nonumber\\
&= -\sqrt{\frac{2}{\pi (\text{tr}A + 2A_{1,2})}} \label{eqn:costf}
\end{align} 

where
\begin{widetext}
\begin{equation}
     A(r, \varepsilon, \varepsilon^{\prime}, N_B^{\prime}, \Delta \phi) := 
     \begin{pmatrix}
      1 + 2 \epsilon^{\prime} (N_B^{\prime} + (1 + N_B^{\prime}) N ) & 
     -2 \sqrt{\epsilon \epsilon^{\prime}} (1 + N_B^{\prime}) \sqrt{N (N + 1)} \cos(\Delta \phi) \\
     -2 \sqrt{\epsilon \epsilon^{\prime}} (1 + N_B^{\prime}) \sqrt{N (N + 1)} \cos(\Delta \phi) &
       1 + 2 \epsilon (1 + N_B^{\prime}) N
     \end{pmatrix}\ 
     \label{eqn:Amtrix}
\end{equation}
\end{widetext}

\begin{figure}[t]
\centering
\includegraphics[scale=0.37]{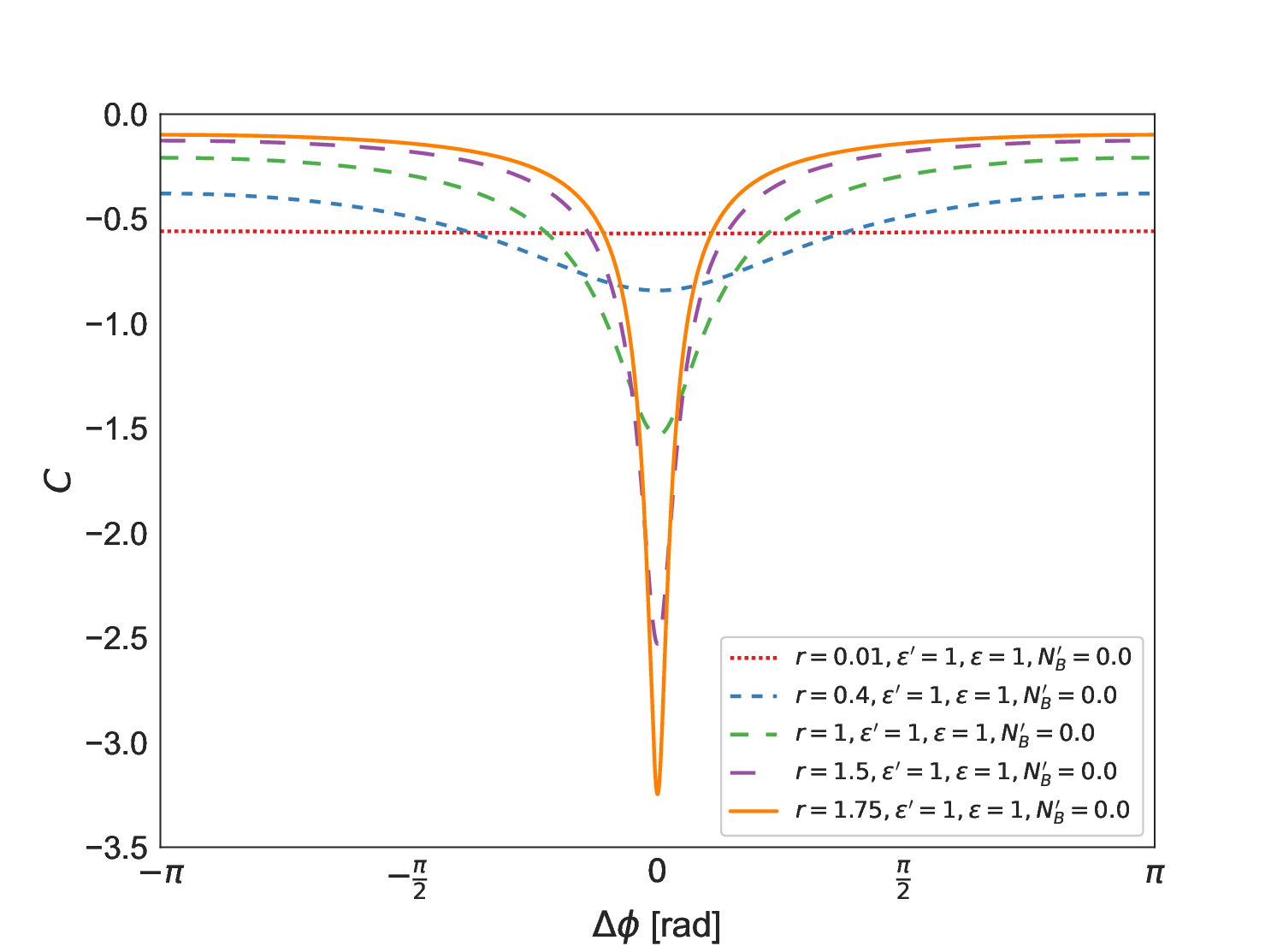}
\caption{\label{fig:2} Cost function landscape for different degrees of squeezing, for the ideal case of no excess noise ($N_B' = 0$) and no losses ($\epsilon^{\prime}=\epsilon^{\prime}=1$), as a function of the phase difference ($\Delta \phi = \phi_c - \phi_0$) between the target phase $U(\phi_0)$ and control phase $V^*(\phi_c)$ unitaries. The cost function directly measures how closely $V^*(\phi_c)$ approximates $U(\phi_0)$. We use the cost to estimate $\Delta \phi$ in the QCA, which serves as a practical gauge of accuracy.}
\end{figure}  
\noindent
is a matrix-valued function, with $A_{1,2}$ and $\text{tr}A$ representing the off-diagonal matrix element and the trace, respectively (see Appendix~\ref{sec:theory}). As we will see later, we find that the theoretical cost function closely approximates the experimentally obtained cost function. 

Figure~\ref{fig:2} shows the behavior of the cost function given in Eq.~(\ref{eqn:costf}) as a function of $\Delta\phi$ for different squeezing parameters, $r$, for the ideal case of zero excess noise, $N^{\prime}_B=0$, and no optical losses, $\varepsilon^{\prime}=\varepsilon=1$ for the probe and conjugate, respectively. Crucially, the landscape of $C$ depends on the squeezing parameter, $r$, with a narrowing of the cost minimum occuring around $\Delta\phi=0$ as $r$ increases.  While the minimum of the cost function occurs at $\Delta\phi=0$, in practice, there will be fluctuations around this minimum, which limit the precision with which the cost can be optimized and thus the precision with which the QCA compiles the parameter(s) of the target unitary. This precision is determined by the width of the cost function. To see why this is the case, one can consider the ideal case of $\varepsilon=\varepsilon ' =1$ and $N_{B}'=0$, i.e., the lossless noise-free case, and proceed by expanding the cost function (normalized by its global minimum value) about $\Delta \phi = 0$. This gives a quadratic scaling of the form
\begin{align}
    &{}-1+{f(N)\Delta\phi^{2} \over 2} + o(\Delta \phi^{2}),
    \label{eqn:narrow} \\
    f(N)&:= \sqrt{N(N+1)}(2\sqrt{N(N+1)} + (2N+1) ),
\end{align}
which implies that success of the QCA (i.e., optimizing to within a neighborhood $\delta$ of the relative minimum value of the cost function) results in learning the phase with an error, as defined by the symbol $\Delta\phi^{2}$, that scales as the reciprocal of the energy squared (proportional to $\delta/N^2$) of the squeezed light. This is similar to the precision obtained from direct phase detection with a Mach-Zehnder interferometer with squeezed light as input to one of its ports~\cite{PhysRevD.23.1693}. 

The general method of utilizing high-probability domains of a homodyne distribution to define a cost function for learning $\Delta \phi$ allows one to obtain an estimate of $\Delta \phi$ with precision that depends on the non-classicality of the probe state. To demonstrate this,
we consider the phase precision achieved by such a learning algorithm when the probing state consists of a pure TMSS operation seeded with $\vert \alpha \rangle \vert \alpha\rangle$,  coherent displacements of amplitude $\alpha\in \mathbb{R}$ in each mode. After the target and variational unitaries are applied on their respective mode, the $(q_{1},q_{2})$ homodyne distribution is then given by
\begin{align}
    P(x_{1},x_{2})&={1\over \pi\sqrt{\det A}}e^{-{1\over \det A}(x-v)^{\intercal}A(x-v)} \nonumber \\
    A&= \begin{pmatrix}
        \cosh 2r & -\sinh(2r)\cos\Delta \phi \\
        -\sinh(2r)\cos\Delta \phi & \cosh 2r
    \end{pmatrix} \nonumber \\
    v&=\begin{pmatrix}
        \sqrt{2}e^{r}\alpha \cos \phi_{c} \\
        \sqrt{2}e^{r}\alpha \cos \phi_{0}
    \end{pmatrix},
\end{align}
where we now consider the total energy per mode to be distributed between the seed and quadrature squeezing according to $N=\alpha^{2}e^{2r} + \sinh^{2}r$. Defining a cost function by $-P(\sqrt{2}e^{r}\alpha,\sqrt{2}e^{r}\alpha)$, i.e., the most likely $(q_{1},q_{2})$-quadrature value occurring for the probe state, followed by normalizing to the global minimum value, gives a function of $\Delta \phi$ of the form
\begin{align}
    &{} -1+2\cosh^{2}r\sinh^{2}r\Delta\phi^{2} \nonumber \\
    &+ O(\alpha^{2})(1+\cosh(4r)+\sinh(4r))\Delta\phi^{4} + o(\Delta \phi^{4}) ,
    \label{eqn:seed}
\end{align}
where only the leading order in $\alpha^{2}$ is kept in the second term. Here the term of order $o(\Delta \phi^{2})$ results from the noise distribution of the TMSS, while the term of order $o(\Delta \phi^{4})$ is a result of the coherent displacement. In the weak-displacement limit ($\alpha^2 \rightarrow 0$), Eq.~(\ref{eqn:seed}) reduces to the same quadratic scaling found in Eq.~(\ref{eqn:narrow}), 
\begin{equation}
-1 + 2N^2\,\Delta\phi^2 + o(\Delta\phi^4)\text{,}
\end{equation}
yielding $\Delta\phi^{2} \propto \delta/N^{2}$ when the cost function is within $\delta$ of the minimum, thereby recovering Heisenberg scaling of the precision, where here $N = \sinh^{2}r$. In the opposite limit of negligible squeezing $r \rightarrow 0$, Eq.~(\ref{eqn:seed}) simplifies to
\begin{equation}
    -1 + {N\over 2}\,\Delta\phi^{4} + o(\Delta\phi^{4}),
    \label{eqn:classical}
\end{equation}
where $N = \alpha^{2}$ represents the energy of the coherent displacement. In this regime, the precision in $\Delta \phi$ obtained by calculating the cost to within $\delta$ of the global minimum scales as $\Delta \phi^2 \propto \sqrt{\delta/N}$, which is actually worse than  standard quantum limit (SQL) scaling with respect to $N$. Achieving SQL scaling in $\Delta \phi$ would require more information from the homodyne distribution. We conclude that on the manifold of seeded TMSSs, the cost function-based readout of the phase difference interpolates between sub-SQL scaling and Heisenberg scaling of precision as the squeezing level increases. Note that the cost function given in Eq.~(\ref{eqn:costf}) is defined in terms of the quadrature noise properties of the resource state and is not applicable to pure coherent states because it contains no information about $\Delta \phi$ in this case.

Although in both Eqs.~(\ref{eqn:narrow}) and (\ref{eqn:seed}) the physical source of the precision advantage (namely quadrature squeezing) is the same as in homodyne readout of a Mach-Zehnder interferometer, the gradient-descent based algorithm learns $\phi_{0}$ by tuning $\phi_{c}$ and calculating the cost function, not by directly estimating  the $\phi_{0}$ phase shift. Our demonstration of this approach in Section \ref{sec:level2} shows that our algorithm is, in principle,  capable of learning unknown target phase shift operations  with precision comparable to Mach-Zehnder interferometry with squeezed light input.

One can additionally see from Fig.~\ref{fig:2} that the expected cost function gradient in the neighborhood of the minimum increases concomitantly with the accuracy of the phase difference estimate as the squeezing parameter $r$ increases, in agreement with the theoretical prediction in ~\cite{tylerUniversalCompilingNo2021}.
It follows that the choice of squeezing parameter, $r$, (or equivalently the energy regime) plays a critical role in the time to solution of our variational QCA. 
For low squeezing values, e.g. $r = 0.01$  shown in red, the cost landscape is relatively flat for all $\Delta \phi$. In this case, given the inevitable reduced precision, as shown by Eq.~(\ref{eqn:narrow}), it will be challenging to discern the cost minimization direction, which will potentially prohibit training. For moderate squeezing values, e.g. $0.4 \leq r \leq 1.5$, the cost function exhibits substantial gradients over all phase difference values, thus making training easier. In the high-squeezing regime $r>1.5$, once within the narrow gorge~\cite{arrasmith2021equivalence}, the sharpness of the cost function will drive the optimization to a faster and more precise final estimate of the target unitary phase.

It is worth noting, however, that in the high-squeezing regime, the sharpness of the cost function makes it such that it approaches a constant exponentially fast for regions of the landscape away from the minimum. Thus, for a random initialization, this energy-dependent exponential flattening of the landscape would, with an exponentially high probability, increase the time to solution of the QCA, eventually precluding optimization altogether as the squeezing parameter $r$ increases~\cite{Volkoff2021,tylerUniversalCompilingNo2021,Zhang_2025}. This is known as the \textit{barren plateau} phenomenon.  

Barren plateaus are a major barrier to the successful scaling of DV variational quantum algorithms~\cite{cerezo2023does}, as their cost landscapes have been shown to flatten exponentially away from minima with the number of qubits~\cite{mcclean2018barren, holmes2021connecting, cerezo2020cost, marrero2020entanglement, holmes2021barren, fontana2023theadjoint, ragone2023unified}. The exponential flattening observed for our CV variational quantum compilation algorithm is crucially different in that it scales with squeezing parameter and not system size. Moreover, the potential offered by tuning the squeezing parameter, $r$, from low to high during training to avoid barren plateaus arguably makes the flattening and sharpening of the landscape with $r$ a feature, rather than a barrier, to training. 

\subsection{\label{sec:level2}Experimental Implementation of a QCA}
To demonstrate a CV quantum compiler based on the configuration shown in Fig.~\ref{fig:1}(a), we use a four-wave mixing (FWM) process in a double-$\Lambda$ configuration in a $^{85}$Rb vapor cell to generate a TMSS~\cite{mccormickStrongRelativeIntensity2007, mccormickStrongLowfrequencyQuantum2008}. In this process two photons from a strong pump beam are absorbed to generate quantum correlated probe ($Pr$) and conjugate ($Co$) beams. As shown in Fig.~\ref{fig:1}(b), the FWM is implemented in a truncated SU(1,1) interferometer~\cite{andersonPhaseSensingStandard2017} configuration with control and target unitaries placed in each of the arms of the interferometers to introduce the control ($\phi_c$) and target ($\phi_0$) phase shifts on the probe and conjugate, respectively. Finally, homodyne detectors, one for each of the modes, are used to measure the quadratures required to evaluate the cost function of the quantum compiler circuit. For these measurements the phase of the respective local oscillator (LO) serves as a phase reference for the phase shift introduced by the control and target unitaries. We generate the required LOs with a second independent FWM process implemented in the same $^{85}$Rb vapor cell~\cite{boyerEntangledImagesFourWave2008}.  

The degree of squeezing of the TMSS, controlled by the squeezing parameter $r$, plays a vital role in ensuring the stability, precision, and efficiency of the quantum compiler. A critical aspect of our experiment is thus maintaining stable and precise control of $r$. To this end, we lock the different parameters of the FWM that determine $r$, specifically the frequency of the laser, the temperature of the $^{85}$Rb vapor cell, and the power of the pump beam. The temperature of the cell and the power of the pump beam are stabilized through proportional-integral temperature and power locks, respectively. We stabilize the laser frequency with a side-band saturation spectroscopy lock using a reference $^{85}\text{Rb}$ vapor cell. The frequency lock allows us to control the one-photon detuning of the pump beam with respect to its corresponding transition. These locks make it possible to keep $r$ constant during each QCA run. Likewise, stable power levels for the TMSS and LOs are essential for a reliable QCA run. We stabilize their powers using proportional-integral power locks applied to the TMSS and LO probe seed beams.

In our experiment, we adjust $r$ by controlling the one-photon detuning of the pump beam. While the second FWM process used to generate the bright LOs is seeded with an input probe beam at a constant power of 145~$\mu$W, this leads to a change in the gain of the FWM process, which in turn changes the powers of the generated LOs and the TMSS. We consider three different values for the squeezing parameter, $r = 0.18$, $0.35$, and $0.74$ (see Section~\ref{sec:levelIII}), to study its impact on our CV QCA. At each degree of squeezing, the optical power of the local oscillator for the probe, $\text{LO}_{Pr}$, is attenuated to match the power of the generated conjugate local oscillator, $\text{LO}_{Co}$, with the power held constant at 17~$\mu$W, 43~$\mu$W, and 330~$\mu$W, respectively, during each QCA run, thereby ensuring that the overall amplification of the homodyne detector remains constant at each degree of squeezing. 

After the FWM, we overlap the probe and conjugate with their respective LO with a polarizing beam splitter (PBS), such that they are orthogonally polarized. We leverage the polarization degree of freedom to implement the unitary optical phase gates, $U(\phi_0)$ and $V^*(\phi_c)$, by transmitting the orthogonally  polarized TMSS and LOs sequentially through a quarter-wave plate (QWP), a half-wave plate (HWP), and another QWP, as seen in Fig.~\ref{fig:1}(b). The initial QWP turns the linear orthogonally-polarized fields into circular orthogonally-polarized fields. The HWP then introduces a relative phase shift between the left and right circularly polarized fields, with the degree of phase shift controlled by rotating the HWP.  Finally, the last QWP converts the circularly polarized light back to linearly polarized light. This process can be visualized by a cyclic path on the Poincar\'{e} sphere that corresponds to the evolution of the light's polarization state~\cite{hariharanGeometricphaseInterferometer, brendelGeometricPhasesTwophoton1995}.  This approach allows us to introduce a well-controlled and stable relative phase shift between the modes of the TMSS and their respective LO.

After interacting with the target and control unitaries, the probe and conjugate are measured with homodyne detection (HD). To implement the HD, the polarization of the probe (conjugate) and its local oscillator $\text{LO}_{Pr}$ ($\text{LO}_{Co}$) are rotated by 45$^\circ$ with an HWP, such that they have orthogonal diagonal polarizations. After the HWP, a PBS is used to combine the fields, with the two outputs sent to a balanced photodetector system. This combination of HWP and PBS, along with the rotation of orthogonally polarized linear fields to orthogonal diagonal polarizations, effectively mimics combining fields with the same linear polarization using a 50/50 beam splitter. 

In order to perform measurements that correspond to the amplitude difference quadrature (as defined by the phase reference of the initial TMSS), the relative phases between the TMSS and LOs need to be locked. To this end, we use a weak coherent state with a constant power of $3~\mu$W to seed to the input probe beam of the FWM process used to generate the TMSS. For this FWM process, the pump power was also held constant at $255~\text{mW}$. To lock the relative phase between the TMSS and the LOs, we take advantage of the unused ports of the PBSs after the FWM process that are used to combine them. These PBSs reflect 1$\%$ of the TMSS and 50$\%$ of the LOs  and are followed by a detection system with a configuration similar to the one used to implement the HD.  

The output of the detection system directly provides an interference pattern that serves as an error signal.  A proportional-integral-differential (PID) controller is then used to lock to the zero crossing of the error signal, which corresponds to locking to the amplitude quadrature, \textcolor{black}{by passing} the control signal to a piezoelectric transducer (PZT) mounted on a mirror in the path of  $\text{LO}_{Pr}$ ($\text{LO}_{Co}$) before the LO is combined with the probe (conjugate) at the PBS, as seen in Fig.~\ref{fig:1}(b). 

It is important to note that the implemented weak displacement of the TMSS by slightly seeding the FWM process does not affect the QCA or the evaluation of the cost function given that the cost function is evaluated based on fluctuations about the mean, as discussed in Section~\ref{sec:level2theory}. For our implementation, the displacement will lead to the maximum of the $X_{-}$ marginal of the Wigner function to be displaced from the origin; however, its profile, which determines the fluctuations about the mean, remains unchanged. 

Once the initial relative phases between the TMSS and LOs are locked, the photocurrents of the measured quadratures are combined with an RF hybrid junction to obtain the difference. The resulting signal is then passed through a  bandpass filter with a full width at half maximum of 500~kHz and centered at 750~kHz to eliminate low-frequency laser noise and the displacement of the Wigner function introduced by weakly seeding the FWM. This is equivalent to performing a measurement with a spectrum analyzer with a 750~kHz analysis frequency and a resolution bandwidth (RBW) of 500~kHz. Finally, the filtered time series data, $X_D(t)$, is digitized with an oscilloscope to accrue statistics to obtain the marginal of the Wigner function required to evaluate the cost function for the quantum compiler.

For the data analysis, the time series quadrature difference data, consisting of $1.2 \times 10^8$ samples, is processed by subtracting the mean voltage difference and scaling the result by the shot noise standard deviation corrected for electronic noise, thereby converting the raw voltage measurements into calibrated quadrature values and yielding a scaled dataset $X_{SD}(t) = X_D(t)/\sqrt{\sigma_{snl}^2 - \sigma_e^2}$, where $\sigma_{snl}^2$ and $\sigma_e^2$ represent the shot noise level and the electronic noise variances, respectively as described in Appendix~\ref{sec:data processing}. To evaluate the cost function, a histogram of the rescaled quadrature data $X_{SD}$ is fitted to a normal distribution to extract its variance, $\sigma_{SD}^2$. Such a fit is justified given the Gaussian nature of the Wigner function. To properly account for the electronic noise, we subtract its contribution (using the same scaling as for the data) in quadrature, such that the variance of the marginal of the Wigner function is given by $\sigma^2 = \sigma_{SD}^2 - \sigma_e^2/(\sigma_{snl}^2 - \sigma_e^2)$. The cost function $C$ can then be easily evaluated by taking the maximum of the resulting normal distribution, which occurs at the origin. A full description of the data processing method is provided in Appendix~\ref{sec:data processing}.

\begin{figure*}[t!]
\includegraphics[scale=.4]{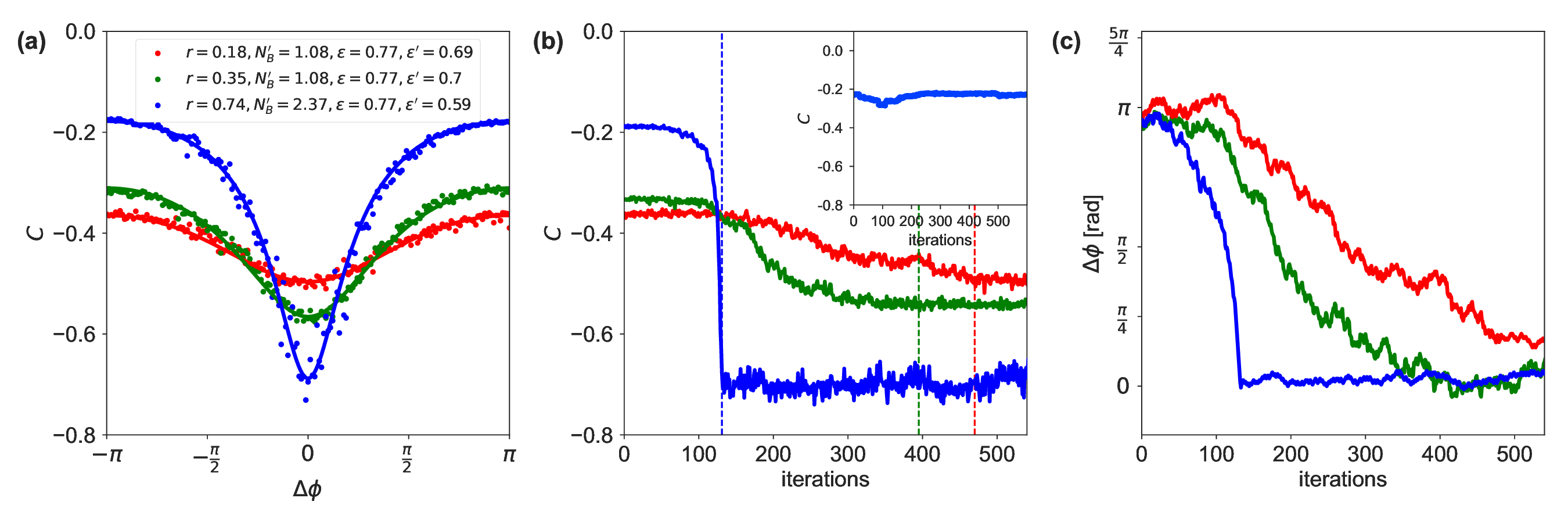}
\caption{\label{fig:3} (a) Theoretical (solid lines) and experimentally measured (dots) cost functions based on the interaction of a TMSS with the phase unitaries  $U = e^{i\phi_0 a^\dagger a}$ and $V^* = e^{-i\phi_c b^\dagger b}$ for different values of $r$. (b) Measured cost function ($C$) as a function of the number of QCA iterations. The dashed lines indicate the iteration at which the QCA has converged for each $r$. The inset shows an instance of a barren plateau phenomenon occurring with the maximum degree of squeezing considered ($r=0.74$). (c)~Estimated phase difference ($\Delta \phi = \phi_c-\phi_0$) as a function of the number of QCA iterations. For all figures red traces correspond to $r=0.18$, green traces to $r=0.35$, and blue traces to $r=0.74$.}
\end{figure*}

\section{\label{sec:levelIII}Results}

In our quantum compiler methodology, we iteratively minimized the cost function derived from the marginal of the Wigner function, as outlined above. During each iteration, which is defined by a window of homodyne detection, we adjusted the phase of the control unitary to minimize the cost function, effectively learning the phase shift implemented by the target unitary. Our results show that in addition to the tunability of the cost function landscape, our QCA platform provides an advantage in the speed and precision needed to efficiently and accurately learn a quantum unitary. 

Figure~\ref{fig:3}(a) shows a comparison between the measured and theoretical cost functions as a function of $\Delta\phi$ for different squeezing parameters, $r$. The data shows the narrowing of the cost function in the neighborhood around the minimum as $r$ increases, which will lead to the precision of the algorithm increasing with $r$. The agreement between the theoretical cost function and the experiment shows that our apparatus was faithful to the theoretical analysis of the cost function. The parameters in the theoretical cost function, given in Eq.~(\ref{eqn:costf}) and Eq.(~\ref{eqn:Amtrix}), were obtained using a maximum likelihood fit of $C$ to the data. We reduce the number of fit parameters by directly measuring the degree of squeezing and losses after the FWM process.  This leaves only two fitting parameters: the excess noise for the input probe seed, $N^{\prime}_B$, and the probe transmissivity, $\varepsilon^{\prime}$. Determining these parameters also allows us to extract the effective value of $r$ for the FWM process. 
The details of our fitting procedure are discussed in detail in Appendix~\ref{sec:parameter_estimation}.

We demonstrated the capabilities of our system by implementing the QCA for three different degrees of squeezing to examine the effect of narrowing the cost function. The QCA results are shown in Figs.~\ref{fig:3}(b) and~\ref{fig:3}(c). For all cases, the same initial value for the phase of the control unitary, $\phi_c$, is used, such that the system is initialized with $\Delta\phi= 3$ rad. We ran the QCA over 550 iterations using gradient descent to determine the next required value of $\phi_c$ to optimize the cost function. In implementing the gradient descent, we used the same training rate for all degrees of squeezing to obtain a fair comparison. When the cost function was minimized to within one standard deviation of its minimum determined by the theoretical fits in Section~\ref{sec:level2theory}, as indicated by the dashed lines in Fig.~\ref{fig:3}(b), the training rate \(\eta\) was reduced by a factor of ten (to \(\eta = 500\)) to improve the stability of the QCA after convergence.
Convergence in gradient-based methods occurs when small gradients near flat regions of a cost landscape near or at its minimum result in negligible parameter updates~\cite{nocedal2006numerical}. In our case, once the cost function is within one standard deviation of the minimum, the stabilization of parameter estimates indicate practical convergence, with further progress limited by numerical precision.

We obtained a metric for the time-to-solution by evaluating the cost function as a function of the number of iterations of the QCA, as shown in Fig.~\ref{fig:3}(b).  We observed a significant reduction in time-to-solution, $t_\mathrm{opt}$, when comparing the QCA runs across increasing degrees of squeezing. Time-to-solution, $t_\mathrm{opt}$, was defined as the number of iterations it took to reach convergence in Fig.~\ref{fig:3}(a) starting from the zero'th iteration. The minimum was empirically determined by taking an average of the cost function values about the center of the cost function where it was flat as shown in~Fig.~\ref{fig:3}(a). We observed a factor of 3.6 speed up in time-to-solution to compile a stable parameterized unitary (i.e. to converge to a stable control phase parameter) measured as the ratio of $t_\mathrm{opt}$ between the $r=0.74$ and $r=0.18$ cases. Furthermore,  the time-to-solution data shown in Table~\ref{tab:precision_main} indicates a nonlinear reduction in $t_\mathrm{opt}$ with respect to the squeezing parameter, tracking with the reduction in quadrature noise.

\begin{table}[ht]
\centering
\begin{tabular}{cccccc}
\toprule
$r$ && $t_{\text{opt}}$ && $\doubleoverline{\Delta \phi}$~[mrad] &  $\sigma_{\overline{\Delta \phi}}$~[mrad] \\
\midrule
0.18 && 470 && 90 & 513.8 \\
0.35 && 395 && 73 & 269.1 \\
0.74 && 130 && 31 & 96 \\
\bottomrule
\end{tabular}
\caption{Measured time-to-solution ($t_{\text{opt}}$), phase difference mean ($\doubleoverline{\Delta \phi}$), and  phase standard deviation in the estimation $\sigma_{\overline{\Delta \phi}}$ for different degrees of squeezing ($r$).}
\label{tab:precision_main}
\end{table}

We next examined how the degree of squeezing influences the precision in estimating the phase parameter of the target unitary. To this end, we ran the QCA multiple times to determine the phase standard deviation in the estimation, defined as the standard deviation, $\sigma_{\overline{\Delta \phi}}$, of the mean phase differences over the $\mathfrak{N}$ runs, such that $\sigma_{\overline{\Delta \phi}}=\sqrt{\overline{\Delta\phi}^{\raisebox{-0.5ex}{\scriptsize 2}}}$. Here $\overline{\Delta\phi}$ is the average over the phase differences within a single run after the QCA has converged. Due to the time it took to implement the 550 iterations of the QCA, which were necessary given the training rate for the gradient descent, we limited the number of runs to $\mathfrak{N} = 5$ for each of the different squeezing parameters. Figure~\ref{fig:3}(c) shows the estimated $\Delta\phi$ as a function of the number of QCA iterations for the different values of the squeezing parameter for one of the runs.

As shown by Eq.~(\ref{eqn:narrow}), the precision in the phase estimate for the compiled unitary, defined as $1/ \sigma_{\overline{\Delta \phi}}$, increased with increasing  the degree of squeezing. To obtain a measure of  $\sigma_{\overline{\Delta \phi}}$, for each of the 5 runs we obtained the mean over the phase differences after the QCA had converged, $\overline{\Delta \phi}$, using  data such as that shown in  Fig.~\ref{fig:3}(c). These $5$ values were then used to calculate the phase difference mean ($\doubleoverline{\Delta \phi}$) and the phase standard deviation, with the values reported in Table~\ref{tab:precision_main}.
As can be seen, the precision increased in a nonlinear fashion consistent with the reciprocal of the reduction in quadrature noise. We show  a factor of 5.4 improvement in the compiled phase precision between the highest and lowest values of $r$.  We validated these results by performing a study using a larger number of runs ($\mathfrak{N}=15$), albeit with a modified training rate for the gradient descent for a faster convergence of the QCA, as described in Appendix~\ref{sec:precision}. As shown in Table~\ref{tab:precision_main}, the phase difference mean decreased as the squeezing parameter, $r$, increased. The increase in $r$ steepened the cost function and yielded larger gradients near the minimum, consistent with previous theoretical studies~\cite{tylerUniversalCompilingNo2021} that indicated that both the precision and the accuracy of the parameter estimation would increase because of the narrowing of the cost landscape~\cite{nocedal2006numerical}. Furthermore, the values obtained for  $\doubleoverline{\Delta \phi}$ demonstrate that our measurements were unbiased as the difference between the expected phase difference mean, $\doubleoverline{\Delta \phi}=0$, and the measured phase difference mean was approximately zero for all measured degrees of squeezing. 

It is important to note that the enhancements obtained for the time-to-solution and precision for the implemented QCA were not dependent on the value of the phase for the target unitary, $\phi_0$ or the initial value used for the phase of the control unitary, $\phi_c$. We showed that is the case by performing QCA runs where i) the target unitary was set to arbitrary phases while the initial control unitary phase was fixed and ii) the control unitary was set to an initial arbitrary phase, while the target unitary phase was fixed (see Appendix~\ref{sec:robustness_qca}).

While for the data shown here the QCA converged for all degrees of squeezing considered, as pointed out in Section \ref{sec:level2theory}, the narrowing of the cost function occurs at the expense of the flattening of the landscape outside the gorge. Such a flattening can lead to barren plateaus, which have been shown theoretically to occur when squeezed resource states are used to define the cost function~\cite{tylerUniversalCompilingNo2021}, a special case of the energy-dependent barren plateau in CV parametrized quantum circuits. Given that our maximum degree of squeezing with $r=0.74$ corresponded to a low energy TMSS resource state, we empirically found that less than $\sim 1\%$ of the time our QCA failed to train completely. We concluded that, in practice, squeezing can be increased to the extent that it provides an enhanced QCA precision while remaining below an acceptable time-to-solution threshold.  An example of one of the rare instances when our QCA failed to train is shown in the inset of Fig.~\ref{fig:3}(b). For QCA runs in which the squeezing parameter was lower than $r = 0.74$ training was always possible. 

The tunability of the cost function landscape of our CV QCA points to the ability to avoid barren plateaus through the implementation of an adaptive QCA in which the compiling is started with a low degree of squeezing to obtain an initial estimate of the parameter of the target unitary. 
Such an estimate would then serve to initialize the phase of the control unitary for a subsequent compiling with a higher degree of squeezing. Compared to random phase initialization for each compiling run, an adaptive approach is capable of counteracting the energy-dependent flattening of the cost landscape, allowing to maximize precision and minimize time-to-solution.

\vspace{0.1 in}
\section{\label{sec:discussion}Discussion}

To the best of our knowledge, our work presents the first experimental implementation of a
squeezing-based variational quantum algorithm for quantum compiling on noisy intermediate-scale quantum (NISQ) hardware in the CV setting. Our proof-of-principle experimental implementation demonstrates the ability of a CV quantum compiler to learn the phase parameter of a target optical phase gate using gradient descent optimization. 

To implement the CV QCA, we leveraged a two-mode squeezed state of light as a quantum resource in our quantum compiler and showed, both theoretically and experimentally, the narrowing of the cost function around $\Delta\phi=0$ as the degree of squeezing increases. We also showed that such a narrowing and associated increased slope led to a reduction in time-to-solution and increased precision in the estimation of the target phase. In particular, we showed that our CV quantum compiler framework gave a $3.6$-fold speed up in the time-to-solution and increased the precision by a factor of $5.4$. This tunability of our CV QCA can enable the implementation of an adaptive QCA to avoid barren plateaus while maximizing precision and minimizing time-to-solution.

An important question is whether these results signify a genuine quantum
advantage over a classical system of equal energy (e.g., a coherent
state). Increasing $r$ raises the total photon number $N = \sinh^2(r)$,
which narrows the neighborhood of the cost-function minimum. Our analysis in
Eq.~(\ref{eqn:narrow}) demonstrates that squeezing dramatically steepens the cost function slope near the neighborhood of the minimum, allowing faster convergence to a more precise phase difference estimate. We also show that our cost function-based readout allows one to approach the Heisenberg scaling of the noise, $\Delta \phi^2 \propto \delta/N^2$, enabled by the two-mode squeezing in our CV quantum compiler. Although losses and
excess noise weaken this ideal behavior, our experimentally observed speed-up
and precision improvement remain robust under realistic conditions. These
findings show that the quantum correlations in the squeezed
resource---not merely the total photon number---drive the enhanced performance,
thereby substantiating a genuine quantum advantage in our CV quantum
compiler.

Our results represent the first step towards the implementation of a universal quantum compiler capable of compiling any CV circuit~\cite{tylerUniversalCompilingNo2021}.  Such a universal quantum compiler  requires an inverse squeezing operation after the target and control unitaries and enables the compilation of all CV unitaries on $M$-modes through a multi-mode SU(1,1) configuration. A non-universal quantum compiler for linear optical CV or, more generally, Gaussian CV circuits, is an important benchmark for optimization of quantum resources on near-term CV hardware, and provides a path forward for the minimization of quantum circuit size and depth in CV quantum algorithms. These developments are critical for the realization of both near-term and fault-tolerant CV quantum computing systems as well as for the characterization of dynamics in quantum systems, such as those of novel quantum materials. 

\section{\label{sec:level1}Acknowledgements}
The authors extend their gratitude for the support provided by the U.S. Department of Energy Office of Science, along with the National Quantum Information Science Research Centers and the Quantum Science Center, under the identifier LA-UR-24-33284.

\appendix

\section{Data Processing}
\label{sec:data processing}
In our experiment we use two balanced homodyne detectors for our measurements to obtain a voltage difference that is proportional to the quadrature difference.  Such quadrature difference measurements are central to phase-sensitive quantum information processing applications based on multimode entangled optical states \cite{10.3788/PI.2022.R06}. In this section, we first derive  the relation between the quadrature difference discussed in Section~\ref{sec:level2theory} and the voltage difference measurements. We then discuss how the data is processed to convert the voltage difference data into quadrature difference data. 

First, we note that the theoretical description of the amplitude quadrature difference $X_{-}$ is of the form 
\begin{align}
    X_{-} & = \frac{x_1 - x_2}{\sqrt{2}}, \\
     & = \frac{v_1 - v_2}{m \sqrt{2}},\
\end{align}
where the amplitude quadratures $x_j$ are related linearly to the voltages measured with the homodyne detectors as $v_j=m x_j$, where $j=1,2$, and $m$ is a proportionality constant. The mean values of the voltage amplitudes ($v_1$ and $v_2$) are the same for each quadrature since the local oscillator powers for the probe and conjugate are made equal to each other in the experiment. To determine $m$, we can consider the variance of the measured voltage in the shot noise limit, such that $\langle \Delta v^2_s \rangle := \langle \Delta v^2_j \rangle = m^2 \langle \Delta x_j^2 \rangle$ regardless of $j$, and note that given our definition of the quadratures the uncertainty principle implies that the variance of the quadratures $x_1$ and $x_2$ for an isotropic minimum uncertainty state is $\langle \Delta x_1^2 \rangle = \langle \Delta x_2^2 \rangle = 1/2$. This makes it possible to determine the constant of proportionality to be $m = \sqrt{2 \langle \Delta v^2_s \rangle}$ for the quadratures of the individual modes in terms of the measured shot noise level. We must take care given that we measure voltage differences in the experiment as opposed to individual beam voltages. Hence, we need to rewrite $m$ in terms of the shot noise limit for the voltage difference, which takes the form $\langle \Delta v^2_{s,-} \rangle = 2 \langle \Delta v_s^2 \rangle$ given that the shot noise voltages from the probe and conjugate beams are uncorrelated, therefore $m = \sqrt{\langle \Delta v^2_{s,-} \rangle}$. The amplitude quadrature difference, $X_{-}$, rewritten in terms of the voltage difference $v_- = v_1-v_2$ and the shot noise measured for the voltage difference, then takes the form \( X_{-} = v_{-}/\sqrt{2 \langle \Delta v_{s,-}^2 \rangle} \), 
such that the variance of the quadrature difference under homodyne detection follows as

\begin{align}
    \langle \Delta X_{-}^2 \rangle = \frac{\langle \Delta v_-^2 \rangle}{ 2 \langle \Delta v^2_{s,-} \rangle }.
     \label{eqn:X_v}
\end{align}

In our experiments, the detection system adds uncorrelated electronic noise to the measurements. Thus, to obtain an accurate measure of the variances of the voltage difference and the shot noise for the voltage difference, we subtract the uncorrelated electronic noise, $\sigma_e^2$,  in quadrature from the measured variance of the squeezed voltage difference, $\sigma_D^2$, and the measured shot noise level for the voltage difference, $\sigma_{snl}^2$. Hence, the variance of the quadrature difference takes the form 
\begin{align}
    \langle \Delta X_{-}^2 \rangle =\frac{1}{ 2} \frac{\sigma^2_D-\sigma^2_e}{ \sigma^2_{snl}-\sigma^2_{e}} = \frac{\sigma^2}{2},
    \label{eqn:y2exp}
\end{align}
where we used
\begin{align}
    \langle \Delta v_-^2 \rangle &= \sigma^2_D-\sigma^2_e \nonumber \\
    \langle \Delta v_{s,-}^2 \rangle &=\sigma^2_{snl}-\sigma^2_{e},
\end{align}
and defined the normalized and corrected variance $\sigma$.

\begin{figure*}[t!]
\includegraphics[scale=.53]{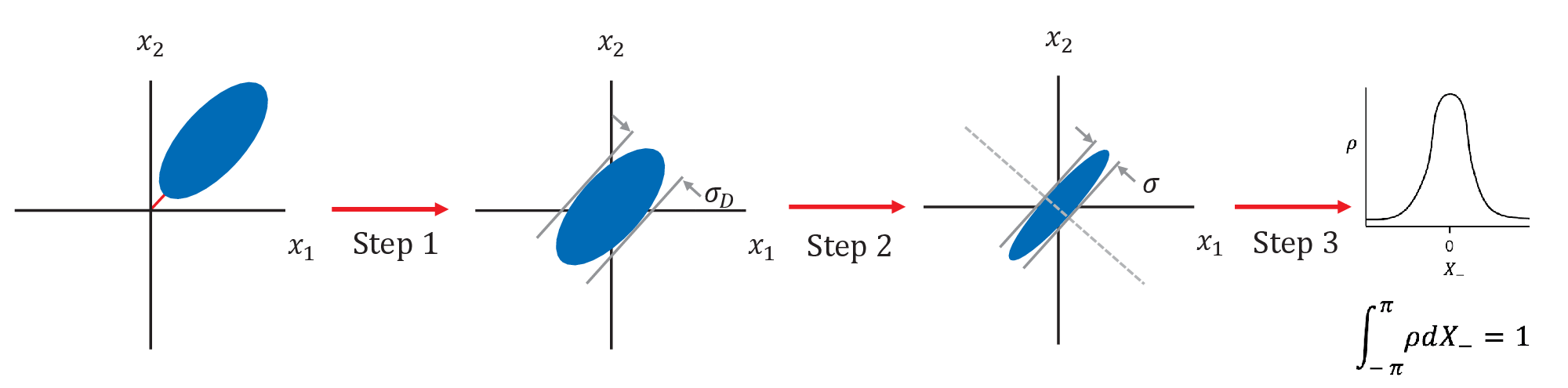}
\caption{\label{fig:1}
Schematic representation of a complimentary data processing procedure used to validate the primary data processing method used. This validation procedure includes three steps: (1) Subtraction of the mean value $\mu_V$ from the data $v_-$ after which the variance, $\sigma_D^2 $, is directly calculated from the time trace, $v_- - \mu_V$. We then correct for the electronic noise giving the electronic noise corrected variance, $\langle \Delta v_-^2 \rangle$. (2) Scaling of $\langle \Delta v_-^2 \rangle$ by the shot noise level of the voltage difference  $\langle \Delta v_{s,-}^2 \rangle$ resulting in the variance, $\sigma^2$, which is then be related related  to  the quadrature difference variance, $\langle \Delta X_{-}^2 \rangle$=$\sigma^2/2$; (3) Finally, we use $\langle \Delta X_{-}^2 \rangle$ to calculate the peak of the homodyne probability density $\rho (X_{-}=0)=1/(2 \pi \langle \Delta X_{-}^2 \rangle)^{\frac{1}{2}}$ for the amplitude quadrature difference. The cost function is  related to this peak value according to $C=-\rho(X_{-}=0)$.} 
\label{fig:data_processing}
\end{figure*}

Having addressed how the quadratures are related to the voltage measurements, we now discuss how we process the data. First, we subtract the mean of the voltage difference $\mu_V$ from the voltage difference data $v_-$ and scale $v_{-}-\mu_V$ by the standard deviation of the homodyne detection voltage difference \(\sqrt{\langle \Delta v^2_{s,-} \rangle}\) (see Eq~(\ref{eqn:X_v}-\ref{eqn:y2exp})). 
 
This yields \(V_{HD} = (v_{-}-\mu_V)/[\sqrt{\sigma_{snl}^2 - \sigma_e^2}]\). Next, we make a  histogram of the scaled time series data $V_{HD}$. The histogram is then fitted to a normal distribution, based on the assumption of a Gaussian distribution in phase space of the quadratures, to determine its variance, \(\sigma_{SD}^2\). We obtain a more accurate estimate of this variance by subtracting the scaled electronic noise variance, to obtain the corrected noise $\sigma^2 = \sigma_{SD}^2 - \sigma_e^2/(\sigma_{snl}^2 - \sigma_e^2)$.
Then, we use Eq.~(\ref{eqn:y2exp}) to obtain the variance of the quadrature difference, $\langle \Delta X_{-}^2 \rangle = \sigma^2/2$.
The final step in the data processing sequence is to determine the peak value of the probability distribution function characterized by $N(0,\langle \Delta X_{-}^2 \rangle)$. This peak corresponds to the maximum of the marginalized Wigner function, $\rho(X_{-}=0)$, as discussed in Section~\ref{sec:level2theory}, such that the cost function $C$ used in our quantum compiler is given by $C =-1/\sqrt{2 \pi \langle \Delta X_{-}^2 \rangle}$. 

To validate our data processing approach, we also implemented a complementary procedure, shown schematically in Fig.~\ref{fig:data_processing}, to the one discussed above. For this complementary procedure, first we subtract the mean voltage difference from the voltage difference data and directly calculate the variance $\sigma_D^2$ of the data $v_{-} - \mu_V$ and make no assumptions about the distribution of the data.  We then compensate for the electronic noise by subtracting its variance, $\sigma_e^2$, from the variance of the voltage difference, $\sigma_D^2$, to obtain the electronic noise corrected variance for the voltage difference $\sigma_D^2 - \sigma_e^2$. Next, we take the ratio of the electronic noise corrected variance of the voltage difference to  the electronic noise corrected variance of the shot noise, resulting in $\sigma^2 = (\sigma_D^2 - \sigma_e^2)/(\sigma_{snl}^2 - \sigma_e^2)$. 
We then use Eq.~(\ref{eqn:y2exp}) to relate this variance to the amplitude quadrature difference variance, such that $\langle \Delta X_{-}^2 \rangle = \sigma^2/2$. Finally, we use this variance, $\langle \Delta X_{-}^2 \rangle$,  to directly evaluate the cost function.
The agreement in $\langle \Delta X_{-}^2 \rangle$ between these two data analysis methods confirms our assumption that the measured quadrature difference is Gaussian.

\section{Precision Data ($\mathfrak{N}=15$)}
\label{sec:precision}
In this section, we validate our observation of the effect the squeezing parameter has on the precision in estimating the parameter $\phi_0$ of the target unitary.  For each squeezing parameter, $r$, we report on an expanded study where $\mathfrak{N}=15$ QCA runs are used to determine the phase difference mean, $\doubleoverline{\Delta \phi}$, and the phase standard deviation,  $\sigma_{\overline{\Delta \phi}}$. To collect this larger data set we increased the learning rate, $\eta$, by a factor of 40 (that is, $\eta=2000$) to expedite the data collection.

\begin{figure}[t]
\includegraphics[scale=0.53]{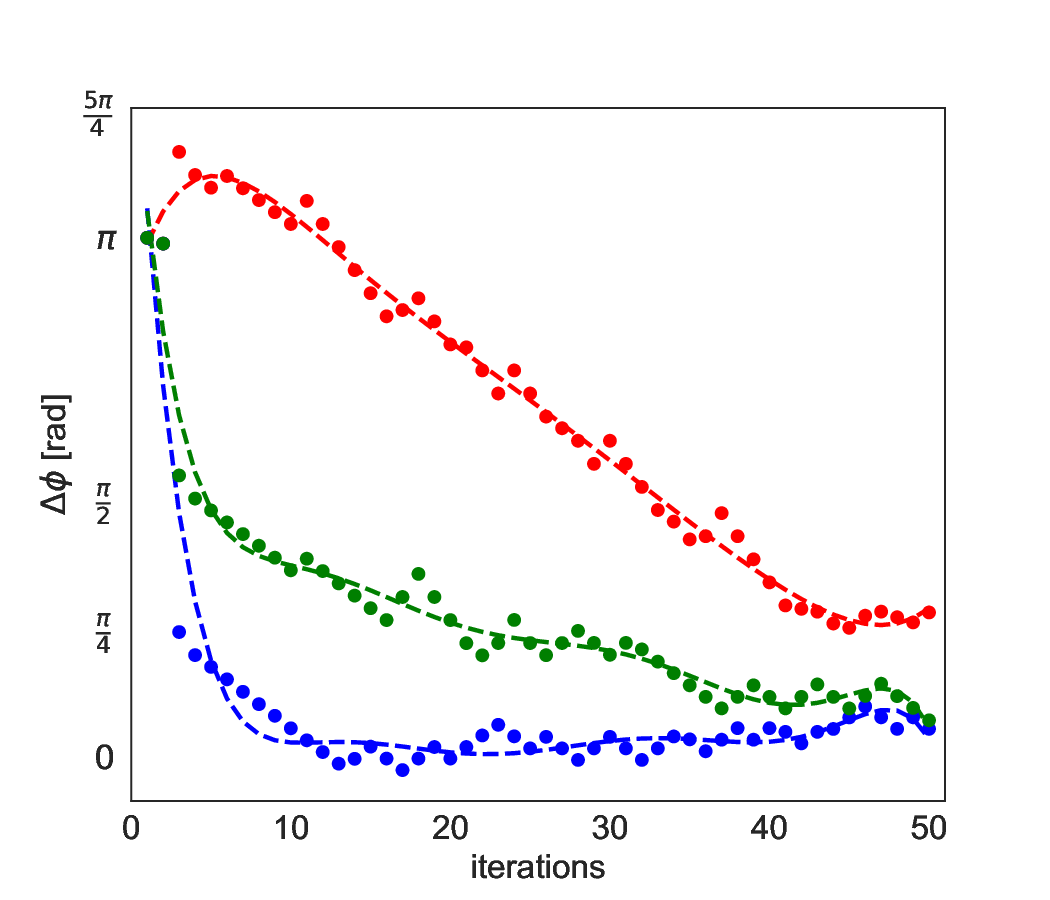} 
\caption{Examples of QCA runs used for precision analysis for different values of $r$. The red, green, and blue data sets (dots) correspond to $r= 0.18$, $0.35$, and $0.74$, respectively. The dashed lines are polynomial fits to guide the eye.}
\label{fig:precision_analysis}
\end{figure}

For each of the 15 QCA runs, we obtain the mean over the phase difference $\Delta \phi$ after the QCA run has converged, $\overline{\Delta \phi}$,  using data like the one shown in Fig.~\ref{fig:precision_analysis}.  We then use these 15 values to obtain the phase difference mean, $\doubleoverline{\Delta \phi}$, and the phase standard deviation, $\sigma_{\overline{\Delta \sigma}}$, which are measures of the accuracy and the uncertainty in the phase difference. As can be seen from Table~\ref{tab:precision_supp}, the phase difference mean decreases as the squeezing parameter, $r$, is increased, consistent with previous numerical simulations~\cite{tylerUniversalCompilingNo2021}. We note that for each $r$ the phase difference mean is higher compared to the study presented in Section~\ref{sec:levelIII}. We attribute this difference to the increased learning rate, as it is known that increasing the learning rate in gradient descent can reduce the accuracy of parameter estimates~\cite{Goodfellow2016,nocedal2006numerical}. Specifically, a higher learning rate causes larger steps during the optimization process, which can lead to overshooting the optimal solution. As a result, larger updates during optimization introduce convergence instabilities, causing a rise in the phase difference mean \doubleoverline{\Delta \phi} and an increase in the phase standard deviation $\sigma_{\overline{\Delta \phi}}$. Consequently, higher learning rates reduce both the overall accuracy and precision of the parameter estimates. In addition, the precision—defined as the reciprocal of the phase standard deviation—scales nonlinearly, tracking with the reciprocal of the degree of squeezing. 

The increase in compiled phase precision between the highest and lowest values of $r$ for the QCA operating at a learning rate $\eta=2000$ was 4.15. This is comparable to the factor of 5.4 improvement in the compiled phase precision measured in Section~\ref{sec:levelIII} where $\eta=50$ was used. These larger data set validate the observations of our precision results discussed in Sections~\ref{sec:levelIII} and ~\ref{sec:discussion}.

\begin{table}[t]
\centering
\begin{tabular}{ccccc}
\toprule
 $r$ & $\doubleoverline{\Delta \phi}$~[mrad] &  $\sigma_{\overline{\Delta \phi}}$~[mrad] \\
\midrule
0.18 & 485.4 & 456.9 \\
0.35 & 125.4 & 241.6 \\
0.74 & 44.6 & 110 \\
\bottomrule
\end{tabular}
\caption{Precision data with $\mathfrak{N}=15$ QCA runs used to calculate the precision and accuracy in compiling the target phase parameter, $\phi_0$. Listed in the table are the phase difference mean, $\doubleoverline{\Delta \phi}$, and the phase standard deviation in the parameter estimation, $\sigma_{\overline{\Delta \phi}}$, for different degrees of squeezing. Here the learning rate used in all QCA runs for these data was $\eta = 2000$.}
\label{tab:precision_supp}
\end{table}

\section{Derivation of the cost function}~\label{sec:theory}
For the task of compiling a target unitary given a discrete or continuous alphabet of gates defining the variational unitary ansatz, a faithful cost function is essential; namely, it must have a unique global minimum when the variational unitary matches the target unitary. When the task is restricted to compiling Gaussian target unitaries with Gaussian variational unitaries and Gaussian resource states, homodyne measurement data can be used to construct a faithful cost function because Gaussian states are equal if and only if the first and second moments of their electromagnetic field quadratures are equal.  Specifically, in the present task of compiling an optical phase shift with a state of nearly zero phase space displacement but anisotropic noise, the second moment of a single phase space marginal of the Wigner function already provides a faithful cost function. Further, cost functions based on only the first and second moments of the homodyne distribution, although not applicable for the task of universal quantum compiling, often have a simple functional form.

The Wigner function of a two-mode CV state $\xi$ is defined as:
\begin{equation}
    W(X) = \int_{\mathbb{R}^{4}} {dX'\over 4\pi^{4}} \, e^{-iX^{\intercal}\Omega X'}\text{tr}\left[ \xi e^{-iX^{'\intercal}\Omega R}\right],
    \label{eqn:wig}
\end{equation}
where $R=(q_{1},q_{2},p_{1},p_{2})^{\intercal}$ is the vector of canonical quadrature operators, $X:=(x_{1},x_{2},y_{1},y_{2})^{\intercal}$, and $\Omega = \begin{pmatrix} 0&I_{2}\\-I_{2}&0\end{pmatrix}$ is the symplectic form on the phase space $\mathbb{R}^{4}$.

The homodyne probability density $P$ of the two $q$-quadratures of $\rho$ is obtained by integrating the Wigner function given in Eq.~(\ref{eqn:wig}) over the $p_{1}$ and $p_{2}$ quadrature values
\begin{equation}
    P(x_{1},x_{2}) := {1\over 4}\int dy_{1}dy_{2}\, W(X).
    \label{eqn:probprob}
\end{equation}
Marginalizing to the difference quadrature $(q_{1}-q_{2})/\sqrt{2}$ produces the univariate probability density
\begin{equation}
    \rho(X_{-}) := \int dX_{+} P\left( {X_{+}+X_{-}\over\sqrt{2}},{X_{+}-X_{-}\over\sqrt{2}} \right),
    \label{eqn:margmarg}
\end{equation}
where $X_{\pm}$ are eigenvalues of $(q_{1}\pm q_{2})\sqrt{2}$. Our non-universal cost function is obtained by finding the maximum of $\rho(X_{-})$, which is simply related to the variance of $X_{-}$. For our optical phase compiling task, we use a model of the resource state $\xi$ that, in addition to being parameterized by the target phase $\phi_{0}$ and the controllable variational phase $\phi_{c}$, takes into account the noisy nature of the four-wave mixing process and attenuation. Specifically,
\begin{widetext}
\begin{equation}
\xi:=   e^{i\phi_c a_1^\dagger a_1} \otimes e^{-i\phi_0 a_2^\dagger a_2} 
    \left( \mathcal{N}_{\epsilon}^{(1)} \otimes \mathcal{N}_{\epsilon'}^{(2)}
    \left[ e^{r(a_1^\dagger a_2^\dagger - \text{h.c.})} \rho_{N_B} \otimes \rho_{N'_B} e^{-r(a_1^\dagger a_2^\dagger - \text{h.c.})} \right] \right)
    e^{-i\phi_c a_1^\dagger a_1} \otimes e^{i\phi_0 a_2^\dagger a_2},
\label{eqn:ststst}
\end{equation}
where $e^{ra^{\dagger}b^{\dagger} - rab}$ is the unitary two mode squeezing operator, $r$ is the squeezing parameter, $N_B$ and $N^{\prime}_B$
are the excess noises of the conjugate and probe, respectively, and 
$\mathcal{N}_{\epsilon}$ and $\mathcal{N}_{\epsilon^{\prime}}$ 
are the noiseless attenuation channels with $\epsilon,\epsilon' \in [0,1]$. 

The lossless approximation can be considered by taking $\epsilon=\epsilon^{\prime}=1$, i.e., $\mathcal{N}_{1}=\mathbf{1}$ is the identity channel.
According to Eq.~(\ref{eqn:probprob}), the homodyne probability density for the quadrature measurements, $q_{1}$ and $q_{2}$, obtained from the Wigner function of Eq.~(\ref{eqn:ststst}), is
\begin{align}
    P(x_{1},x_{2})&= \langle q_{1}=x_{1}\vert \langle q_{2}=x_{2}\vert \xi \vert q_{1}=x_{1}\rangle \vert q_{2}=x_{2}\rangle \nonumber \\
    &= {1\over \pi \sqrt{f(\phi_{0},\phi_{c})}}e^{-{1\over f(\phi_{0},\phi_{c})}x^{T}A(\phi_{0},\phi_{c})x},
\end{align}
where $x=(x_{1},x_{2})^{\intercal}$,
\begin{align}
    f(\phi_0,\phi_c)&= \left( 1 + 2\epsilon \left( N_B + (1 + N_B + N'_B) N_S \right) \right) 
    \left( 1 + 2\epsilon' \left( N'_B + (1 + N_B + N'_B) N_S \right) \right) \\
    &- 4 \epsilon \epsilon' \left( 1 + N_B + N'_B \right)^2 N_S (N_S + 1) \cos^2(\phi_c - \phi_0),\nonumber
\end{align}
and
\begin{equation}
    A(\phi_0, \phi_c) = 
    \begin{pmatrix}
        1 + 2\epsilon' \left( N'_B + (1 + N_B + N'_B) N_S \right) & -2\sqrt{\epsilon \epsilon'} (1 + N_B + N'_B) \sqrt{N_S (N_S + 1)} \cos(\phi_c - \phi_0) \\
        -2\sqrt{\epsilon \epsilon'} (1 + N_B + N'_B) \sqrt{N_S (N_S + 1)} \cos(\phi_c - \phi_0) & 1 + 2\epsilon \left( N_B + (1 + N_B + N'_B) N_S \right)
    \end{pmatrix}.
\end{equation}

$A$ is a matrix-valued function that simplifies to the following when the four-wave mixing process is seeded by a noisy probe and a noisless conjugate ($N_{B}=0$)
\begin{equation}
  A(r, \varepsilon, \varepsilon^{\prime},N_B^\prime,\Delta \phi) := 
     \begin{pmatrix}
      1 + 2 \epsilon^{\prime} (N_B^{\prime} + (1 + N_B^{\prime}) N ) & 
     -2 \sqrt{\epsilon \epsilon^{\prime}} (1 + N_B^{\prime}) \sqrt{N (N + 1)} \cos(\Delta \phi) \\
     -2 \sqrt{\epsilon \epsilon^{\prime}} (1 + N_B^{\prime}) \sqrt{N (N + 1)} \cos(\Delta \phi) &
       1 + 2 \epsilon (1 + N^{\prime}_B) N
     \end{pmatrix}\text{,} 
\end{equation}
where $\Delta\phi:= \phi_{c}-\phi_{0}$. We can then change variables by the $SO(2)$ transformation
\begin{align}
X_{+}&= {x_{1}+x_{2}\over\sqrt{2}}\nonumber \\
X_{-}&= {x_{1}-x_{2} \over \sqrt{2}}
\label{eqn:change1}
\end{align}
to get the marginal density for $X_{-}$.  The transformation  to the $(X_{+},X_{-})$ variables takes the form
\begin{footnotesize}
\begin{align}
P\left( {X_{+}+X_{-}\over\sqrt{2}},{X_{+}-X_{-}\over\sqrt{2}} \right) &= {1\over \pi \sqrt{f(\phi_{0},\phi_{c})}}e^{-{1\over 2f(\phi_{0},\phi_{c})}\begin{pmatrix} X_{+}&X_{-}\end{pmatrix}\begin{pmatrix} 1& 1\\ 1 & -1 \end{pmatrix}A(\phi_{0},\phi_{c})\begin{pmatrix} 1& 1\\ 1 & -1 \end{pmatrix}\begin{pmatrix} X_{+}\\X_{-}\end{pmatrix}} \nonumber \\
&= {1\over \pi \sqrt{f(\phi_{0},\phi_{c})}}e^{-{1\over 2f(\phi_{0},\phi_{c})}\begin{pmatrix} X_{+}&X_{-}\end{pmatrix}\begin{pmatrix} \text{tr}A+2A_{1,2} & 0 \\ 0 & \text{tr}A-2A_{1,2}\end{pmatrix}\begin{pmatrix} X_{+}\\X_{-}\end{pmatrix}}, 
\label{eqn:change2}
\end{align}
\end{footnotesize} 
where $A_{1,2}$ is the $1,2$ matrix element of the matrix-valued function $A$ and $\text{tr}A$ is its trace. 
Doing the Gaussian integral over variable $X_{+}$ gives the marginal probability density according to Eq.~(\ref{eqn:margmarg})
\begin{equation}
\rho(X_{-},r,\epsilon,\epsilon^{\prime}, N^{\prime}_B,\Delta\phi)= \sqrt{2 \over \pi (\text{tr}A+2A_{1,2})} e^{ -{\text{tr}A-2A_{1,2}\over 2f(\phi_{0},\phi_{c})} X_{-}^{2}}\text{.}
\label{eqn:change3}
\end{equation}
This Gaussian function (Eq.~\ref{eqn:change3}) has a global maximum at $X_{-}=0$ with variance $(\text{trA}+\text{A}_{\text{1,2}})/4$. Therefore, the non-universal cost function takes the form $C(r,\epsilon,\epsilon^{\prime}, N^{\prime}_B,\Delta\phi):= -\rho(0,r,\epsilon,\epsilon^{\prime}, N^{\prime}_B,\Delta\phi)$.
\end{widetext}

\section{Parameter Estimation from Experimental Data}
\label{sec:parameter_estimation}

In this section we describe our method for estimating the parameters of the cost function (squeezing parameter \( r \), transmission efficiencies \( \epsilon^{\prime} \) for the probe and \( \epsilon \) for the conjugate, and excess noise \( N^{\prime}_B \) of the seed for the four-wave mixing (FWM) process) based on measurements of the experimental apparatus and twin beams. These measurements are then used to constrain the fit of the experimentally measured cost function to the theory and reduce the number of free parameters. Without constraining these parameters the model becomes non-identifiable. 

First, since the conjugate beam is far from atomic resonance, it does not experience absorption losses as it propagates through the $^{85}$Rb vapor cell. Thus, \( \epsilon \) is determined by  the detection efficiency (optical losses and photodiode quantum efficiency) after the vapor cell, which we measure to be \( \epsilon = 0.77 \). 
 
This leaves three free parameters: \( r \), \( \epsilon^{\prime} \), and \( N^{\prime}_B \). However, certain combinations of these parameters can still produce indistinguishable fits to our measurements. This phenomenon, known as non-identifiability, prevents the extraction of unique estimates for \(r\), \(\epsilon'\), and \(N'_B\) from the measured data. In our model, this non-identifiability arises from the interdependence between these parameter in the FWM process.
 
Our goal then is to extract the parameter values for $r$ and $\epsilon^{\prime}$ that are intrinsic to the FWM process based on measured values of the amplitude quadrature noise, probe excess noise, $N^{\prime}_B$, and losses. The measured values will be affected by these two parameters . To this end, we introduce a model to derive an analytical relation for the squeezing parameter $r$ in terms of $\epsilon^{\prime}$ and  $N^{\prime}_B$, which allows us to reduce  the number of free parameters in our fitting procedure to two ($\epsilon^{\prime}$ and $N^{\prime}_B$) and address the issue of non-idenitfiability. 
We start with the expression for the amplitude difference variance $\langle \Delta X_{{-}}^2 \rangle$ as described in Section~\ref{sec:theory}, that is
\begin{align}
    \langle \Delta X_{{-}}^2 \rangle &= \frac{1}{4}(\text{trA}+2\text{A}_{\text{1,2}})\\
    &= 1 + \epsilon^{\prime} N^{\prime}_B + (\epsilon + \epsilon^{\prime})(N^{\prime}_B + 1)\sinh^2 r \nonumber \\ 
    &\quad - \sqrt{\epsilon \epsilon^{\prime}}(N^{\prime}_B+1)\sinh 2r \cos(\Delta \phi)\\
    &= \frac{1}{2} \bigg[ \left(\epsilon' N_{\text{in}} + \frac{1}{2} \epsilon\right) \cosh^2(r) + \frac{1}{2} (1 - \epsilon) \nonumber \\
    &\quad + \left(\epsilon N_{\text{in}} + \frac{1}{2} \epsilon'\right) \sinh^2(r) + \frac{1}{2} (1 - \epsilon') \nonumber \\
    &\quad - \sqrt{\epsilon \epsilon'} (2 N_{\text{in}} + 1) \sinh(r) \cosh(r) \cos(\Delta \phi) \bigg] \text{,} \label{eq:variance_X_minus}
\end{align}

where $N_{in}$ is the noise of the seed to the FWM process and is given by the sum of the shot noise and the excess noise $N^{\prime}_B$ used for the theory model: $N_{in}=N^{\prime}_B+1/2$. In the experiment, the amplitude difference squeezing is optimal at $\Delta \phi=0$.  Thus, we can set $\Delta \phi=0$ in Eq.~(\ref{eq:variance_X_minus}) to relate the amplitude quadrature difference noise in our model to the minimum amplitude quadrature difference noise we measured $\langle \Delta X_{{-},\text{m}}^2 \rangle$. This allows us to recast  Eq.~(\ref{eq:variance_X_minus}) as a quadratic equation of the form $AR^2+BR+C=0$ with coefficients

\begin{align}
    A &= \frac{(\epsilon' + \epsilon) N_{\text{in}}}{2} + \frac{\epsilon' + \epsilon}{4} + \frac{\sqrt{\epsilon \epsilon'} (2 N_{\text{in}} + 1)}{2} \\
    B &= \frac{(\epsilon' - \epsilon) N_{\text{in}}}{2} + \frac{\epsilon - \epsilon'}{4} + \frac{2 - \epsilon' - \epsilon}{2} - 2\langle \Delta X_{{-,m}}^2 \rangle \\
    C &= \frac{(\epsilon' + \epsilon) N_{\text{in}}}{8} + \frac{(\epsilon' + \epsilon)}{16} - \frac{\sqrt{\epsilon \epsilon'} (2 N_{\text{in}} + 1)}{8}\text{,}
\end{align}

where $R=\text{exp}(-2r)/2$ is the noise level of the amplitude quadrature difference for the ideal case of no losses or excess noise. From here we obtain the analytical expression for $r$ as a function of $\epsilon'$, $N'_B$, and the measured value of squeezing to constrain the parameter space when fitting to the theory.

\begin{figure}[h!]
    \centering
    \includegraphics[width=0.5\textwidth]{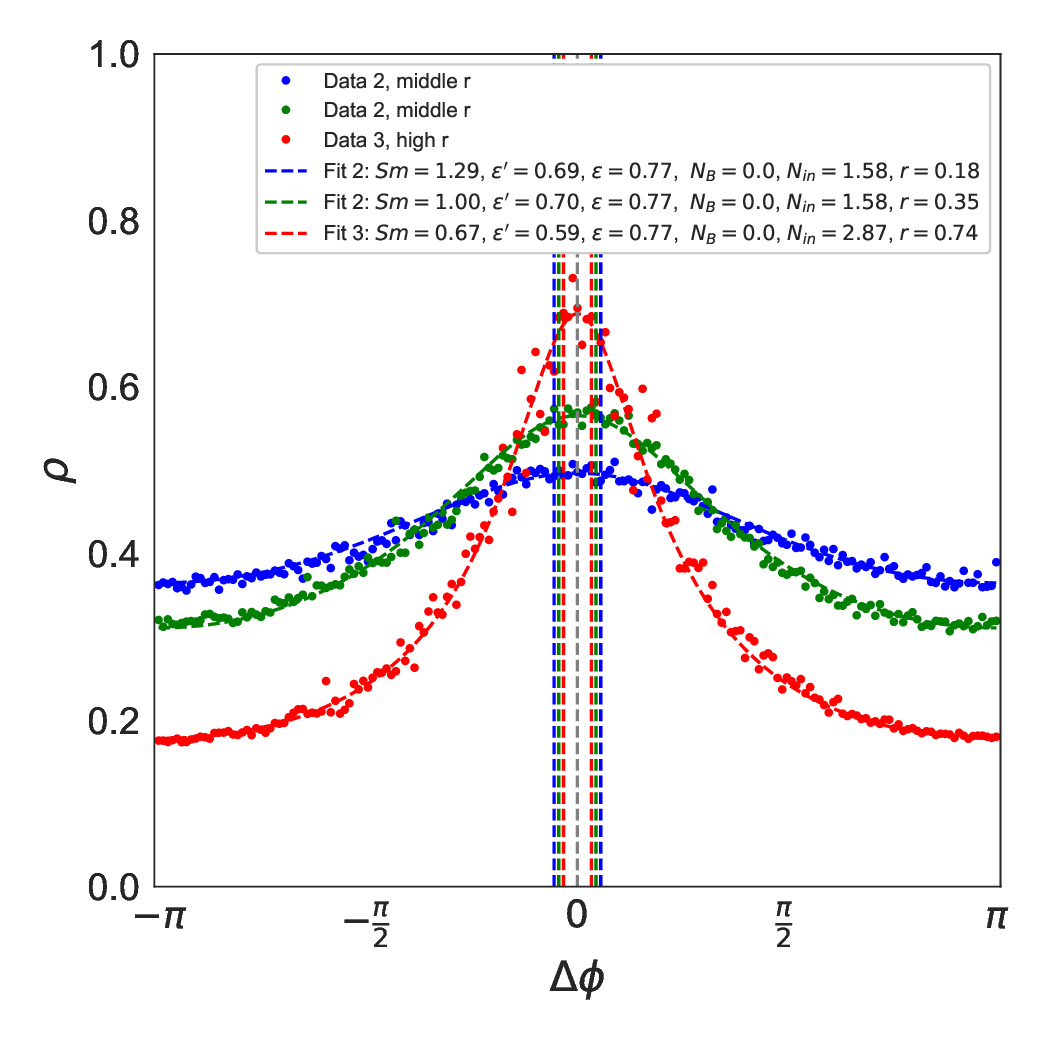}
    \caption{ Comparison of experimental data (dots) and theoretical fits (dashed curves) for the marginal probability density $\rho$. The measured amplitude difference noise, $\langle\Delta X_{-,m}^2\rangle$, used in the fitting procedure, is determined by averaging the measured amplitude quadrature noise data within a range encompassing $12\%$ of the full width at half maximum (FWHM) around the peak of the theoretical $\rho$. The $12\%$ FWHM bounds are indicated by the dashed vertical lines for each data set.}
    \label{fig:fit_parameter_val}
\end{figure}

We use this model in our fitting procedure as follows. We first solve for the quadratic equation  to relate $r$ in our fits to the other two free parameters and constrain its value with the measured amplitude quadrature difference. To obtain an unbiased estimates of the experimental amplitude quadrature difference $\langle \Delta X_{{-,m}}^2 \rangle$, we take an average of the data points about $\Delta \phi = 0$ that lie within $12\%$ of the full width at half maximum (FWHM) of the cost function for the experimental data sets, as shown by the vertical colored dashed lines in Fig.~\ref{fig:fit_parameter_val}. Here $12\%$ of the FWHM was chosen as it minimized the residuals in our fits to the theory model. We then express the cost function $C = -[2\pi \langle \Delta X_{{-}}^2 \rangle]^{-\frac{1}{2}}$ in terms of the free parameters \( \epsilon^{\prime} \), \( N^{\prime}_B \) and the constrained $r( \epsilon^{\prime}, N^{\prime}_B)$, and 
performed a nonlinear least squares minimization to fit our theoretical cost function to the data. This minimization is equivalent to maximum likelihood since the data is normally distributed. In the least squares optimization the range of the two free parameters was bounded as shown in Table~\ref{tab:parameter_values}. The lower bound for the input noise and the upper bound for the probe transmissivity were determined through measurements of the experimental apparatus. These bounds on $N_{in}$ and $\epsilon^{\prime}$ along with the functional relation $r(N_{in},\epsilon^{\prime})$ sufficiently constrained the fits, enabling consistent parameter estimates to be determined. 

\begin{table}[h!]
    \centering
    \begin{tabular}{ccc}
        \hline
         \text{$N_{in}$ bounds} & \text{$\epsilon^{\prime}$ bounds} \\ \hline
         \{0.71, 10\} & \{0, 0.7\} \\ \hline
         \{0.71, 10\} & \{0, 0.7\} \\ \hline
        \{0.71, 10\} & \{0, 0.6\} \\ \hline
    \end{tabular}
    \caption{Parameter ranges for $\epsilon^{\prime}$ and $N^{\prime}_B$ used in the maximum liklehood optimization of the cost function $C$.}
    \label{tab:parameter_values}
\end{table}

The transmissivity of the probe through the atomic medium was measured for each degree of squeezing, with the laser  locked at each of the corresponding one-photon detunings and the pump blocked, to obtain an estimate of atomic absorption. Once the pump is turned on for the FWM, we expect the atomic absorption to increase due the opical pumping.  Thus, the ratio of the optical power before and after the cell was taken as the upper bound for the contribution of the atomic absorption to the probe transmissivity for each case. The total probe tranmissivity ($\epsilon^{\prime}$) is then taken to be the product of the transmission through the $^{85}$Rb vapor and the detection transmissivity ($\epsilon$). 

The input noise $N_{in}$ of the seed probe was measured through direct detection while tuning the laser off resonance with the cell at room temperature, with no pump beam, and at the same power used for the experiments. While the measured input noise was $N_{in}=0.71$, the estimated $N_{in}$ parameter from the fits exceeds the measured value. This results in part from the FWM process, which leads to excess noise due to the distributed nature of loss and gain in medium that leads to amplification of uncorrelated noise from the losses. Our model does not incorporate such distributed gain and loss dynamics to describe the squeezing parameter and the absorption loss of the FWM process~\cite{mccormick2008strong, jasperse2011relative}. This simplification underestimates the effective $N_{in}$ that takes into account the input excess probe noise and the excess noise from the FWM process.  However, qualitatively our model exhibits the physical properties we expect from our system: the excess noise and the loss in the probe both increase with increasing squeezing parameter. This is a result of a larger FWM gain and the use of detuning to increase $r$, with smaller one-photon detuning for larger $r$~\cite{turnbull2013role}. 

A model in which noise is coupled into the measurement due to an imperfect homodyne detection was also evaluated. Such a model would take into account the excess noise that couples from the spatially multimode squeezed light when the probe and conjugate homodyne detectors are not mode matched to the corresponding quantum correlated spatial modes~\cite{Gupta:20}. This model showed similar agreement in goodness-of-fit with the experimental data as the one used in Section~\ref{sec:levelIII}. For the noisy homodyne model, the squeezing parameter $r$ was constrained in the same manner as previously described, leaving three free parameters: the probe transmissivity and the two noise terms corresponding to the noisy homodyne detection for the probe and conjugate. To compare the models, we used the Akaike Information Criterion (AIC)~\cite{Usami2003,Yano2023,Lougovski2009,Yin2011,Vadiraj2021,vanEnk2013,Liu2016}, a metric for model selection that balances model fit and complexity, expressed as, expressed as
\begin{equation}
    \text{AIC} = 2k - 2 \log(L_{\text{max}}),
\end{equation}
where \(k\) is the number of model parameters and \(\log(L_{\text{max}})\) is the log of the maximum likelihood. Since the noise of the likelihood function for both models is described by Gaussian statistics, we may write the maximum log-likelihood as
\begin{equation}
    \log(L_{\text{max}}) = -\frac{n}{2} \left( 1 + \log(2\pi) + \log\left(\frac{\text{RSS}}{n}\right) \right).
\end{equation}
Here, \(n\) is the number of data points, and \(\text{RSS}\) denotes the residual sum of squares of the fit. The noisy seed model yielded the lowest AIC value, with a difference of 10 compared to the noisy homodyne model. This indicates that the noisy homodyne model is significantly less supported by the data, suggesting that its additional free parameter does not sufficiently improve the model fit and may lead to overfitting~\cite{Burnham2002}. Given this result, combined with the direct measurement of excess noise in the seed used in the experiment, we adopted the noisy seed model for analysis in the Section~\ref{sec:levelIII}.

\section{Robustness of QCA for Arbitrary Target and Control Phases}
\label{sec:robustness_qca}
\begin{figure}[b]
\centering
\includegraphics[width=0.41\textwidth]{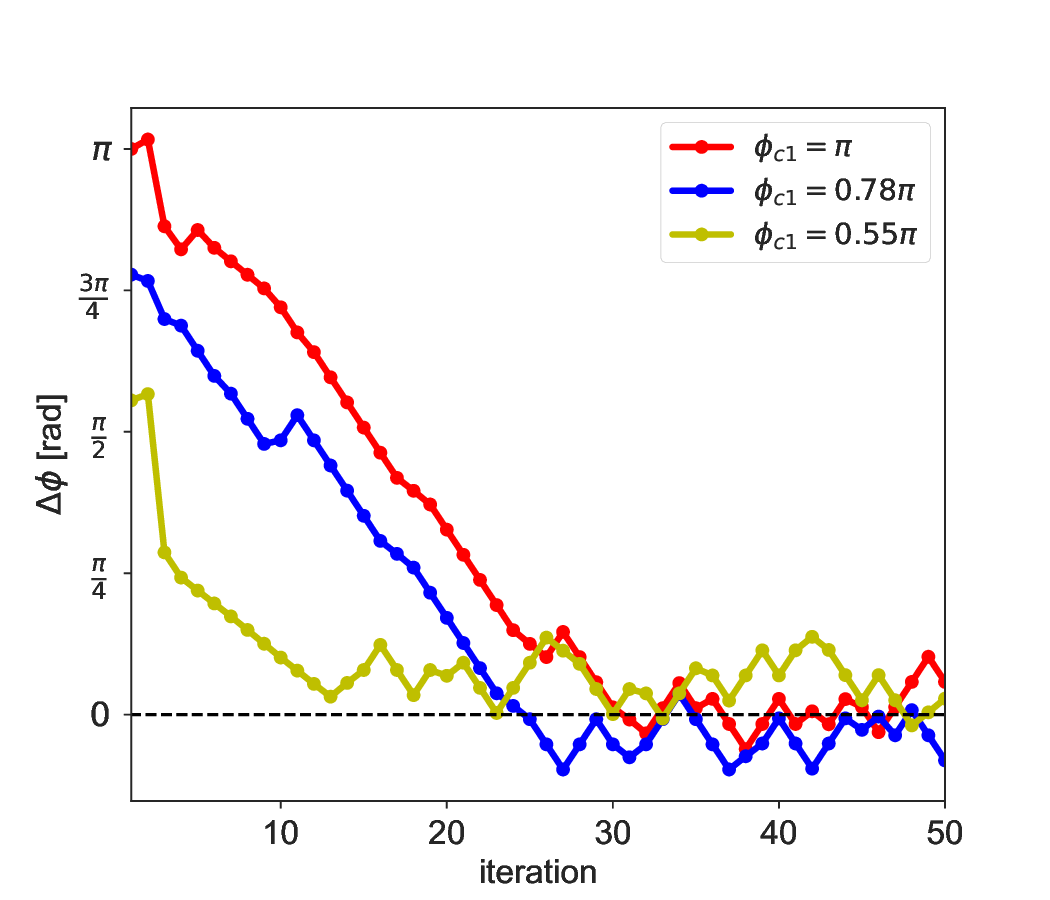} 
\caption{Convergence of the QCA algorithm for different initial control phases $\phi_c$ with the target phase fixed to $\phi_0=0$.}
\label{fig:target_phase_fixed_init_control_phase_varied}
\end{figure}

The robustness of the QCA algorithm to the target and initial control phases is crucial in quantum compiling to ensure the compiler performs consistently.  To evaluate the consistency of our quantum compiler, we ran the QCA for two cases as follows:  in case a) we adjusted  the initial control phase for each run once to either $\pi\text{, } 0.78\pi \text{, or } 0.55\pi$ while keeping the target phase fixed at \(\phi_0 = 0\), and in case b) we adjusted the target phase for each run to either $0.49\pi\text{, } 0.33\pi \text{, or } 0$ while keeping the initial control phase fixed at \(\phi_c = \pi\). For both cases, the QCA used a fixed squeezing parameter of $r = 0.74$ and the same learning rate as in Appendix~\ref{sec:precision} ($\eta = 2000$). The algorithm's ability to converge under these conditions was then evaluated. 

Convergence in gradient descent algorithms, including the QCA algorithm, can often be determined by monitoring the change in the value of the parameters to be optimized. When the parameters stop changing significantly between iterations, the algorithm has likely reached a stable point, as extensively discussed in the optimization literature~\cite{nocedal2006numerical}. In our experiments, this heuristic approach aligns with the observed behavior of the QCA algorithm, where a reduction in the parameter change correlates with convergence to a stable phase difference.

\begin{figure}
\centering
\includegraphics[width=0.41\textwidth]{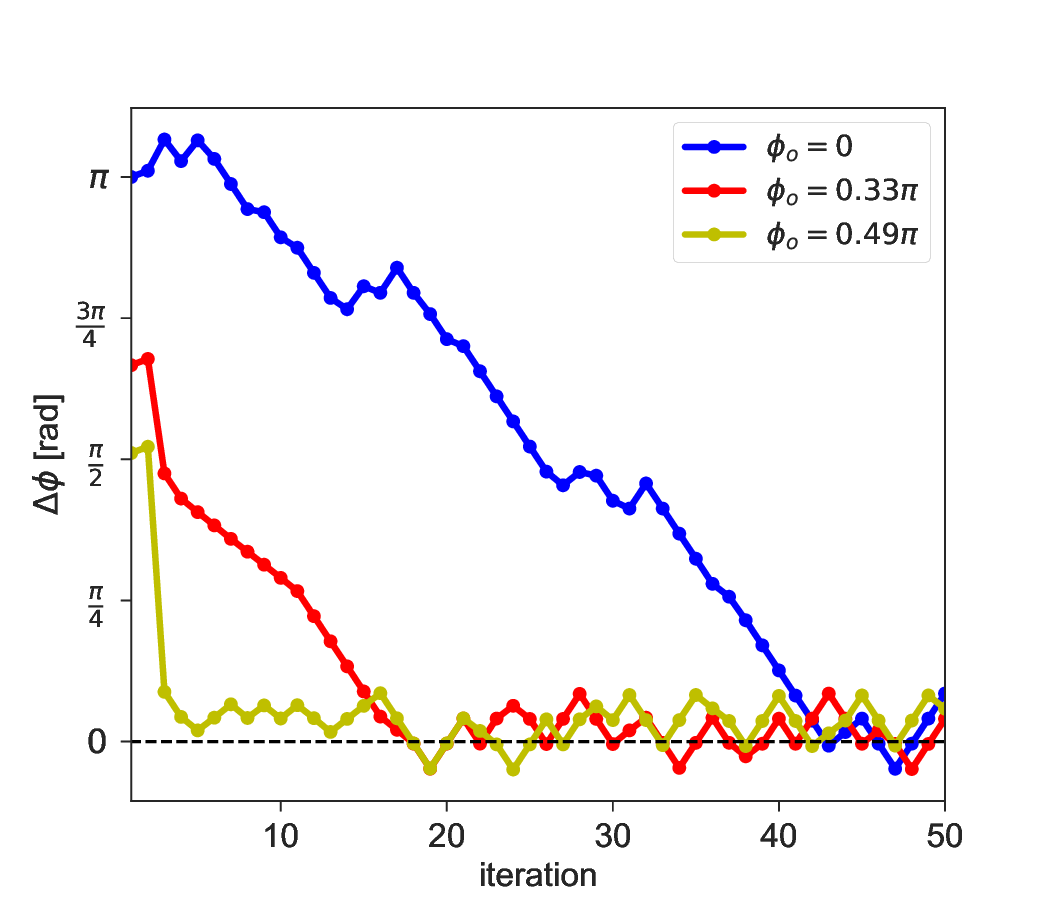} 
\caption{Convergence behavior of the QCA algorithm with the initial control phase set to $\phi_c=\pi$ for different values of the target phase $\phi_0$.}
\label{fig:init_control_phase_fixed_target_phase_varied}
\end{figure}
\vspace{2 in}
In case a), we kept the target phase fixed at $\phi_0=0$ and adjusted the initial control phase parameter to three distinct values: $\phi_c = \pi$, $ 0.78 \pi$, and $0.55 \pi$ radians. The results, shown in Fig.~\ref{fig:target_phase_fixed_init_control_phase_varied}, show that the QCA algorithm reliably converges to the correct phase independent of the initial control phase. For these data, over the thee runs, we calculated the phase difference mean $\doubleoverline{\Delta \phi} = 36$~mrad and the phase standard deviation $\sigma_{\overline{\Delta \phi}} = 128$~mrad, which tracks with the results in Table~\ref{tab:precision_supp} for the highest squeezing case. Likewise, for case b) we fix the initial control phase to $\phi_c=\pi$ and adjusted the target phase parameter to three distinct values: $\phi_0 = 0$, $0.33 \pi$, and $0.49 \pi$ radians. The results, shown in Figure \ref{fig:init_control_phase_fixed_target_phase_varied}, show that the QCA algorithm reliably converges to the correct target phase. For these data, over the three runs, we calculated the phase difference mean $\doubleoverline{\Delta \phi} = 22$~mrad and the phase standard deviation $\sigma_{\overline\Delta \phi} = 78$~mrad, which also tracks with the results in Table~\ref{tab:precision_supp} for the highest squeezing case.


\end{document}